\documentclass[reprint,
preprintnumbers,
nofootinbib,
amsmath,amssymb,
aps,
prd,
superscriptaddress,
]{revtex4-2}

\usepackage{graphicx}
\usepackage{dcolumn}
\usepackage{siunitx}
\usepackage{bm}
\usepackage[hidelinks]{hyperref}
\usepackage[all]{hypcap}
\usepackage{xcolor}
\usepackage{enumitem}
\usepackage{multirow}
\usepackage{orcidlink}

\hypersetup{
	colorlinks,
	linkcolor={red!50!black},
	citecolor={blue!50!black},
	urlcolor={blue!80!black}
}

\DeclareMathOperator{\Tr}{Tr}

\begin{document}

\preprint{ADP25-26/T1288}

\title{Novel insight into centre-vortex geometry in four dimensions}

\author{Jackson A. Mickley \orcidlink{0000-0001-5294-2823}}
\affiliation{Centre for the Subatomic Structure of Matter, Department of Physics, The University of Adelaide, South Australia 5005, Australia}

\author{Chris Allton \orcidlink{0000-0003-0795-124X}}
\affiliation{Department of Physics, Swansea University, Swansea, SA2 8PP, United Kingdom}
\affiliation{School of Mathematics and Physics, The University of Queensland, St.\ Lucia, Brisbane, Queensland 4072, Australia}

\author{Ryan Bignell \orcidlink{0000-0001-8401-1345}}
\affiliation{School of Mathematics and Hamilton Mathematics Institute, Trinity College, Dublin 2, Ireland}

\author{Derek B. Leinweber \orcidlink{0000-0002-4745-6027}}
\affiliation{Centre for the Subatomic Structure of Matter, Department of Physics, The University of Adelaide, South Australia 5005, Australia}

\begin{abstract}
Centre-vortex surfaces are mapped out in four dimensions within the framework of \(\mathrm{SU}(3)\) lattice gauge theory to understand the role of secondary loops that develop in three-dimensional visualisations of centre-vortex structure, appearing separate from the percolating cluster. Loops that initially appear disconnected in three-dimensional slices can originate from the same connected surface in four dimensions depending on the surface's curvature. For the first time, these secondary loops are identified as ``connected" or ``disconnected" with respect to the vortex sheet, allowing new insight into the evolution of centre-vortex geometry through the finite-temperature phase transition. At low temperatures, we find that secondary loops of any length primarily lie in the same sheet percolating the four-dimensional volume. Only a handful of small secondary sheets disconnected from the percolating sheet are identified. Above the phase transition, the vortex structure is still found to be dominated by a single large sheet but one that has aligned with the temporal dimension. With the near\linebreak absence of any curvature orthogonal to the temporal dimension, connected secondary loops become vanishingly rare. Other novel quantities, such as the four-dimensional density of secondary sheets and the sheet sizes themselves, are analysed to build a complete picture of centre-vortex geometry in four dimensions.
\end{abstract}

\maketitle

\section{Introduction} \label{sec:intro}
The nature of quark confinement is one of the longest-standing questions in quantum chromodynamics (QCD). In the context of static heavy quarks, confinement is typically inferred through an area-law falloff for large Wilson loops \(\mathcal{C}\) \cite{Wilson:1974sk},
\begin{equation}
	\langle W(\mathcal{C}) \rangle \sim \exp\left(-\sigma A(\mathcal{C})\right) \,,
\end{equation}
where \(\sigma\) is the string tension. In the presence of light quarks, this simple picture breaks down due to string breaking at large separations. Instead, the absence of a K{\"a}llen-Lehmann representation of the gluon propagator implies that the corresponding physical states are confined. This is signalled by positivity violation in the Schwinger function at large Euclidean times \cite{Aubin:2003ih, Bowman:2007du}.

One aspect of vacuum structure that has been well established as a leading prospect for a confinement mechanism is the presence of centre vortices in the ground-state fields \cite{tHooft:1977nqb, tHooft:1979rtg, Nielsen:1979xu, Greensite:2003bk, Greensite:2016pfc}. These form closed two-dimensional sheets in four dimensions, and have been studied extensively in the framework of \(\mathrm{SU}(N)\) lattice gauge theory \cite{DelDebbio:1996lih, DelDebbio:1997ep, Langfeld:1997jx, Kovacs:1998xm, DelDebbio:1998luz, Engelhardt:1998wu, Langfeld:1998cz, Faber:1998qn, deForcrand:1999our, Bertle:1999tw, Engelhardt:1999fd, Faber:1999sq, Engelhardt:1999wr, Kovacs:2000sy, deForcrand:2000pg, Langfeld:2001cz, Langfeld:2003ev, Engelhardt:2003wm, Gattnar:2004gx, Bornyakov:2007fz, Bowman:2008qd, Bowman:2010zr, OMalley:2011aa, Trewartha:2015nna, Trewartha:2017ive, Spengler:2018dxt, Biddle:2019gke, Biddle:2022zgw, Biddle:2022acd, Biddle:2023lod, Mickley:2024zyg, Mickley:2024vkm, Mickley:2025ksp}. The removal of centre vortices has been demonstrated to result in a vanishing string tension, indicating the loss of an area law \cite{deForcrand:1999our, Langfeld:2003ev, Biddle:2022zgw}. In addition, centre-vortex removal with dynamical fermions leaves no sign of positivity violation in the Schwinger function \cite{Biddle:2022acd}. These findings highlight the importance of centre vortices in understanding the nonperturbative phenomena of QCD.

Centre vortices that ``percolate" spacetime, i.e.\ span the four-dimensional volume, naturally generate an area law for Wilson loops in those orientations pierced by vortices \cite{Engelhardt:1998wu}. This is the observed behaviour for both space-space and space-time orientations at low temperatures. However, \(\mathrm{SU}(N)\) Yang-Mills theory is known to undergo a phase transition at a critical temperature \(T_d\) beyond which confinement breaks down. This is reflected in the centre-vortex structure through an alignment of the vortex sheet along the temporal dimension \cite{Langfeld:1998cz, Bertle:1999tw, Engelhardt:1999fd, Mickley:2024zyg}. This results specifically in the near absence of space-time Wilson loops pierced by a vortex. Meanwhile, centre vortices still pierce and generate an area law for space-space Wilson-loop orientations above \(T_d\).

These characteristics have previously been visualised by mapping out the centre-vortex clusters that form in three-dimensional slices of the full four-dimensional lattice \cite{Biddle:2019gke, Biddle:2023lod, Mickley:2024zyg, Mickley:2024vkm}. Here, the closed two-dimensional sheets that exist in four dimensions are reduced to closed one-dimensional loops. To be precise in our terminology, we use ``cluster" for any generic connected group of vortices. Accordingly, ``sheets" are clusters in four dimensions, while ``loops" are clusters in three-dimensional slices.

\begin{figure}
	\includegraphics[width=\linewidth]{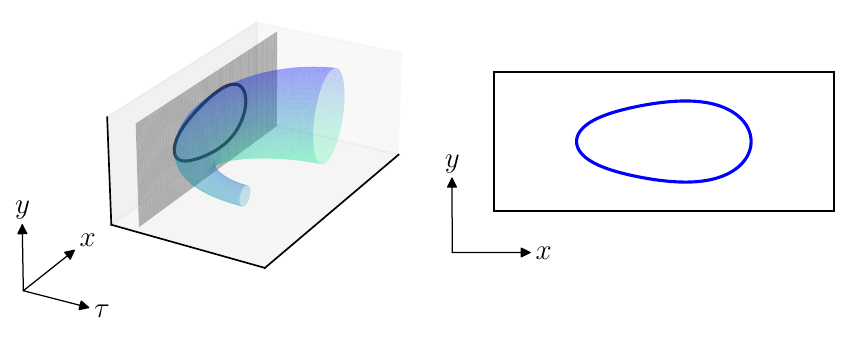}
	\includegraphics[width=\linewidth]{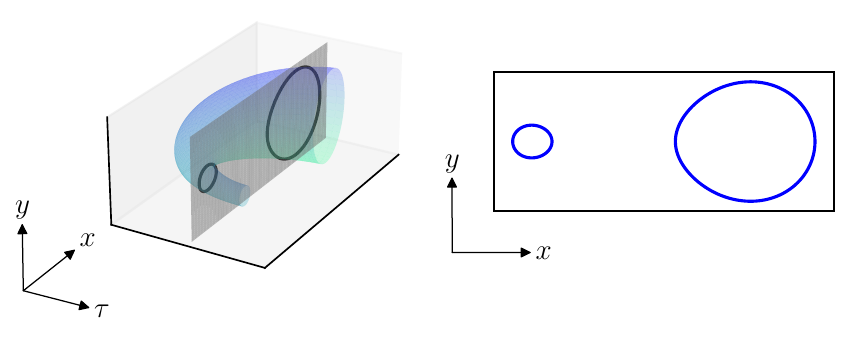}
	\caption{\label{fig:disconnectedloops} Adapted from Ref.~\cite{Mickley:2024vkm}. An illustration of how disconnected loops can arise by slicing through a single curved surface. Here, the \(\tau\) coordinate is being held fixed such that we are looking at an \(xy\) cross section. Depending on the slice coordinate, we can see either a single closed curve (\textbf{top}) or two disconnected curves (\textbf{bottom}).}
\end{figure}

Though some aspects of centre-vortex geometry can be deduced in this manner, other features are lost in slicing through the vortex sheet that exists in four dimensions. For example, it has previously been discussed how vortices that \emph{appear} disconnected in three-dimensional slices can lie in the same connected sheet in four dimensions, due to the sheet's curvature \cite{Mickley:2024vkm}. An illustration of this idea is provided in Fig.~\ref{fig:disconnectedloops}.
This can also be seen with a sufficiently fine lattice spacing---in an animation over three-dimensional slices, one can observe such loops walking across the lattice and joining together \cite{Mickley:2025ksp}.

Due to the evolution in centre-vortex geometry across the phase transition, one would expect to find a substantial change in the connectedness (or otherwise) of these secondary loops in four dimensions. Nevertheless, it is in general not possible to ascertain whether two loops are connected in four dimensions purely by looking at the three-dimensional structure.

It is thus prescient to detect the distinct centre-vortex surfaces that form in the full four-dimensional lattice. This has previously been carried out in \(\mathrm{SU}(2)\) and exploited to study topological aspects of the vortex sheets, such as their orientability and Euler characteristics \cite{Bertle:1999tw}. Our emphasis will instead be placed on the physical ramifications of vortex connectedness in four dimensions with regard to the deconfinement phase transition.

We perform this analysis in pure \(\mathrm{SU}(3)\) gauge theory at a range of temperatures that span the phase transition, building on the work in Ref.~\cite{Mickley:2024zyg} that studied vortex geometry via three-dimensional slices. In addition to the four-dimensional connectedness of loops discussed above, we will investigate other novel quantities made available through this complete picture of centre-vortex geometry. These include the proportion of vortices that lie in the largest sheet, along with the density and sizes of any secondary sheets. This will build a deeper understanding of the relationship between centre vortices and confinement.

This paper is structured as follows. In Sec.~\ref{sec:centrevortices}, centre vortices and their identification on the lattice are succ\-inctly reviewed. Our four-dimensional analysis of centre-vortex geometry is detailed in Sec.~\ref{sec:4D}, with accompanying visualisations that incorporate the updated cluster information. Comparisons are drawn to our prior work that restricted cluster identification to three-dimensional slices. Section \ref{sec:analysis} presents our detailed analysis of the various four-dimensional vortex properties new to this work. Finally, we conclude our main findings in Sec.~\ref{sec:conclusion}.

\section{Centre vortices} \label{sec:centrevortices}
Centre vortices \cite{tHooft:1977nqb, tHooft:1979rtg, Nielsen:1979xu} are regions of the gauge field that carry magnetic flux quantised by the centre of SU(3),
\begin{equation}
	\mathbb{Z}_3 = \left\{ \exp\left(\frac{2\pi i}{3}\, n \right) \mathbb{I} \;\middle|\; n = -1,0,1 \right\} \,.
\end{equation}
Physical vortices in the QCD ground-state fields have a finite thickness. Any Wilson loop that encircles a vortex acquires a factor of an element of \(\mathbb{Z}_3\),
\begin{equation}
	W(\mathcal{C}) \to z\,W(\mathcal{C}) \,.
\end{equation}
On the lattice, thin centre vortices are extracted through a well-known gauge-fixing procedure that seeks to bring each link variable \(U_\mu(x)\) as close as possible to an element of \(\mathbb{Z}_3\), known as maximal centre gauge (MCG), followed by projecting onto this nearest \(\mathbb{Z}_3\) element. The thin vortices that emerge comprise closed surfaces, or ``sheets'', in four-dimensional Euclidean spacetime, and therefore one-dimensional structures in a three-dimensional slice of the four-dimensional spacetime.

Fixing to MCG is performed by finding the gauge transformation \(\Omega(x)\) to maximise the functional \cite{Montero:1999by}
\begin{equation} \label{eq:mcg}
	R = \sum_{x,\,\mu} \,\left| \Tr U_\mu^{\Omega}(x) \right|^2 \,.
\end{equation}
The links are subsequently projected onto the centre,
\begin{equation}
	U_\mu^{\Omega}(x) \to Z_\mu(x) = \exp\left(\frac{2\pi i}{3} \, n_\mu(x) \right) \mathbb{I} \in \mathbb{Z}_3 \,,
\end{equation}
with \(n_\mu(x) \in \{-1,0,1\}\) identified as the centre phase nearest to \(\arg \Tr U_\mu(x)\) for each link. Finally, the locations of vortices are identified by nontrivial plaquettes in the centre-projected field,
\begin{equation} \label{eq:centreprojplaq}
	P_{\mu\nu}(x) = \prod_\square Z_\mu(x) = \exp\left(\frac{2\pi i}{3} \, m_{\mu\nu}(x) \right)\mathbb{I} \,,
\end{equation}
with \(m_{\mu\nu}(x) = \pm 1\). The value of \(m_{\mu\nu}(x)\) is referred to as the ``centre charge", and we say the plaquette is pierced by a vortex.

Although gauge dependent, numerical evidence indicates that the projected vortex locations correspond to the physical ``guiding centres" of thick vortices in the original fields \cite{DelDebbio:1998luz, Montero:1999by, Faber:1999gu, Langfeld:2003ev}. This allows the significance of centre vortices to be investigated through the vortex-only field \(Z_\mu(x)\).

The ensembles employed in this work are the same as in Ref.~\cite{Mickley:2024zyg}. They use the Iwasaki renormalisation-group-improved action \cite{Iwasaki:1983iya, Iwasaki:1996sn} with \(\beta = 2.58\), corresponding to a lattice spacing of \(a \simeq 0.1\)\thinspace fm \cite{CP-PACS:2004das}. Their spatial volume is \(32^3\) and the temporal extent is varied to obtain three temperatures below \(T_d\) and three above \(T_d\). Their details are reproduced in Table \ref{tab:ensembles} for convenience.
\begin{table}
	\centering
	\caption{\label{tab:ensembles} The number of sites \(N_\tau\) in the temporal dimension for our ensembles, with the corresponding temperatures both in terms of the critical deconfinement temperature \(T_d\) and in MeV. The conversion is performed using \(r_0 \, T_d \simeq 0.746\) \cite{Francis:2015lha} and the Sommer scale \(r_0 \simeq 0.5\)\thinspace fm \cite{Sommer:1993ce}.}
	\begin{ruledtabular}
		\begin{tabular}{S[table-format=2.0]S[table-format=1.2]S[table-format=3.0]}
			\multicolumn{1}{c}{\(N_\tau\)} & \multicolumn{1}{c}{\(T/T_d\)} & \multicolumn{1}{c}{\(T\) (MeV)} \\
			\colrule
			64 & 0.10 &  30 \\
			12 & 0.55 & 162 \\
			8 & 0.83 & 243 \\
			6 & 1.10 & 324 \\
			5 & 1.32 & 389 \\
			4 & 1.65 & 486
		\end{tabular}
	\end{ruledtabular}
\end{table}

\section{Centre-vortex structures in four dimensions} \label{sec:4D}
To understand our approach for identifying the distinct centre-vortex sheets in four dimensions, we first recall how clusters can be identified in three-dimensional slices. Our four-dimensional algorithm is then a natural extension of this procedure.

\begin{figure}
	\centering
	\includegraphics[width=\linewidth]{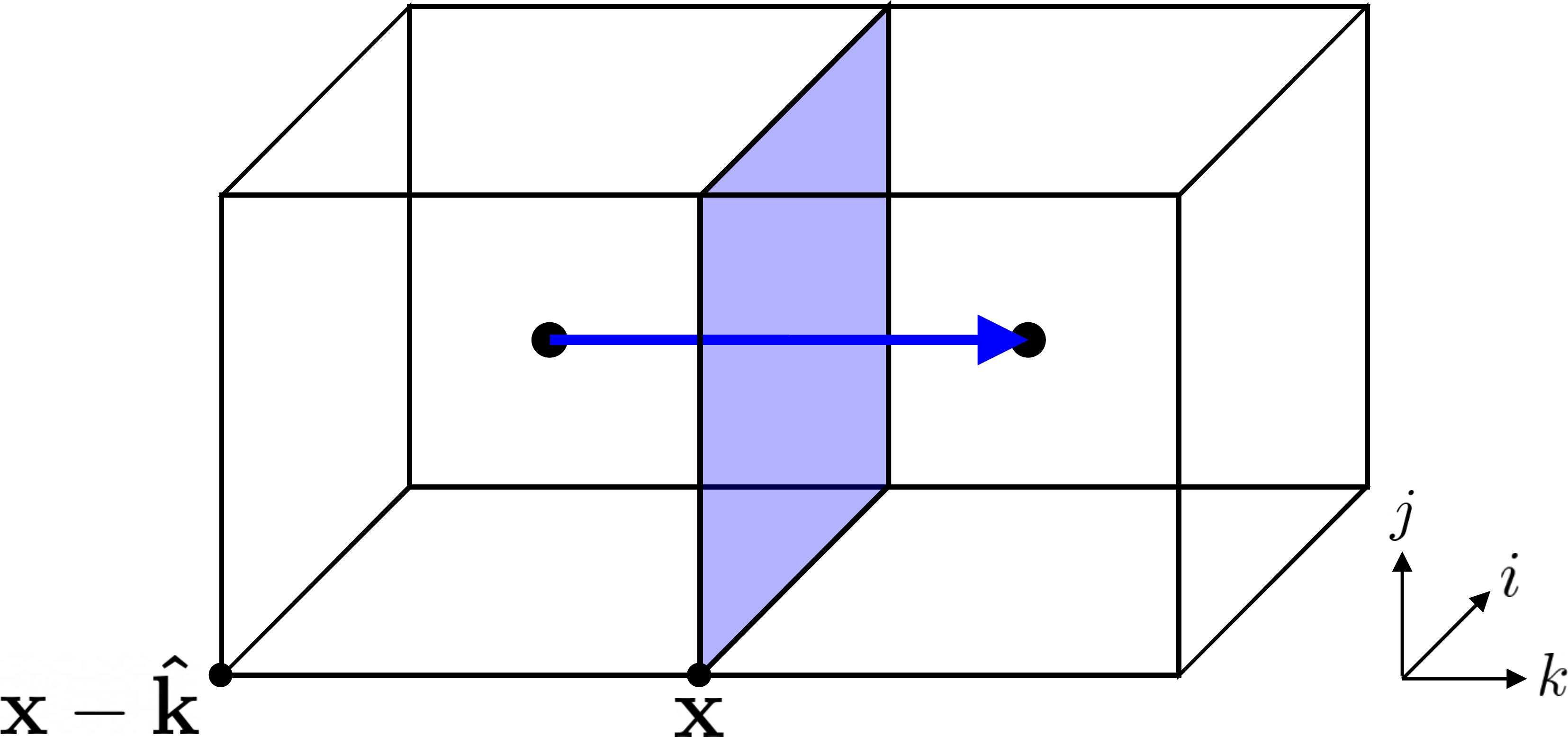}
	\caption{\label{fig:connectingplaquettes} Illustration for identifying connecting plaquettes in three-dimensional slices of the lattice. The middle \(ij\) plaquette at \(\mathbf{x}\) is pierced by an \(m = +1\) vortex, rendered as a unit arrow on the dual lattice pointing in the \(+\mathbf{\hat{k}}\) direction and piercing the plaquette. Due to conservation of centre charge, there must be at least one vortex connection to the left, and at least one to the right. Consequently, one can exclusively check the faces of the elementary cubes at \(\mathbf{x}\) and \(\mathbf{x - \hat{k}}\) to identify the next nontrivial plaquettes. This holds also if \(m = -1\) and the vortex were to point in the \(-\mathbf{\hat{k}}\) direction. In four dimensions, one need simply repeat this procedure for the two dimensions orthogonal to the pierced plaquette.}
\end{figure}

Centre-vortex clusters constitute closed loops in three-dimensional slices. The vortices exist on the dual lattice, piercing the associated nontrivial plaquettes. Therefore, if an \(ij\) plaquette at \(\mathbf{x}\) is pierced by a vortex, the possible connecting plaquettes form nothing but the faces of the elementary cubes at \(\mathbf{x}\) and \(\mathbf{x - \hat{k}}\) (\(\epsilon_{ijk} = 1\)). This idea is illustrated in Fig.~\ref{fig:connectingplaquettes}.
There is a total of 10 plaquettes to check for vortices at each step, 5 faces for each elementary cube. One can thus map out a loop by selecting an initial pierced plaquette, identifying the connecting nontrivial plaquettes by checking the faces of these two elementary cubes, and then iteratively applying this procedure to each of those plaquettes in turn. The process would terminate once all connecting nontrivial plaquettes in a given step have already been visited.

Moving to four dimensions, there are now two orthogonal directions \(\kappa\), \(\lambda\) for a given \(\mu\nu\) plaquette (\(\epsilon_{\mu\nu\kappa\lambda} = 1\)). Noting that the arrows plotted on the dual lattice (as in Fig.~\ref{fig:connectingplaquettes}) lie within the sheet, here the connecting nontrivial plaquettes can be identified by implementing the above three-dimensional procedure independently for \(\kappa\) and \(\lambda\). First, \(\lambda\) is held fixed and one imagines the elementary cubes formed by translating the \(\mu\nu\) plaquette one step forward and backward in the \(\kappa\) direction. The plaquettes that form these cubes are checked for vortices. Then, the roles of \(\kappa\) and \(\lambda\) are interchanged---\(\kappa\) is held fixed, and one checks the forward and backward elementary cubes in the \(\lambda\) direction. After the faces of all four elementary cubes have been visited, all possible connecting plaquettes have been accounted for. This can then be applied recursively to map out centre-vortex sheets in four dimensions.

One of our primary goals is to understand the extent to which vortex loops that appear disconnected in three-dimensional slices are connected in four dimensions. To achieve this, we produce visualisations with two different colour schemes. In one, the vortices will be coloured by the loop they belong to in the three-dimensional slice. This is how the visualisations in Ref.~\cite{Mickley:2024zyg} were produced. In the second scheme, they will be coloured by the sheet they belong to in four dimensions. The two schemes can subsequently be compared.

\begin{figure}
	\centering
	\includegraphics[width=\linewidth]{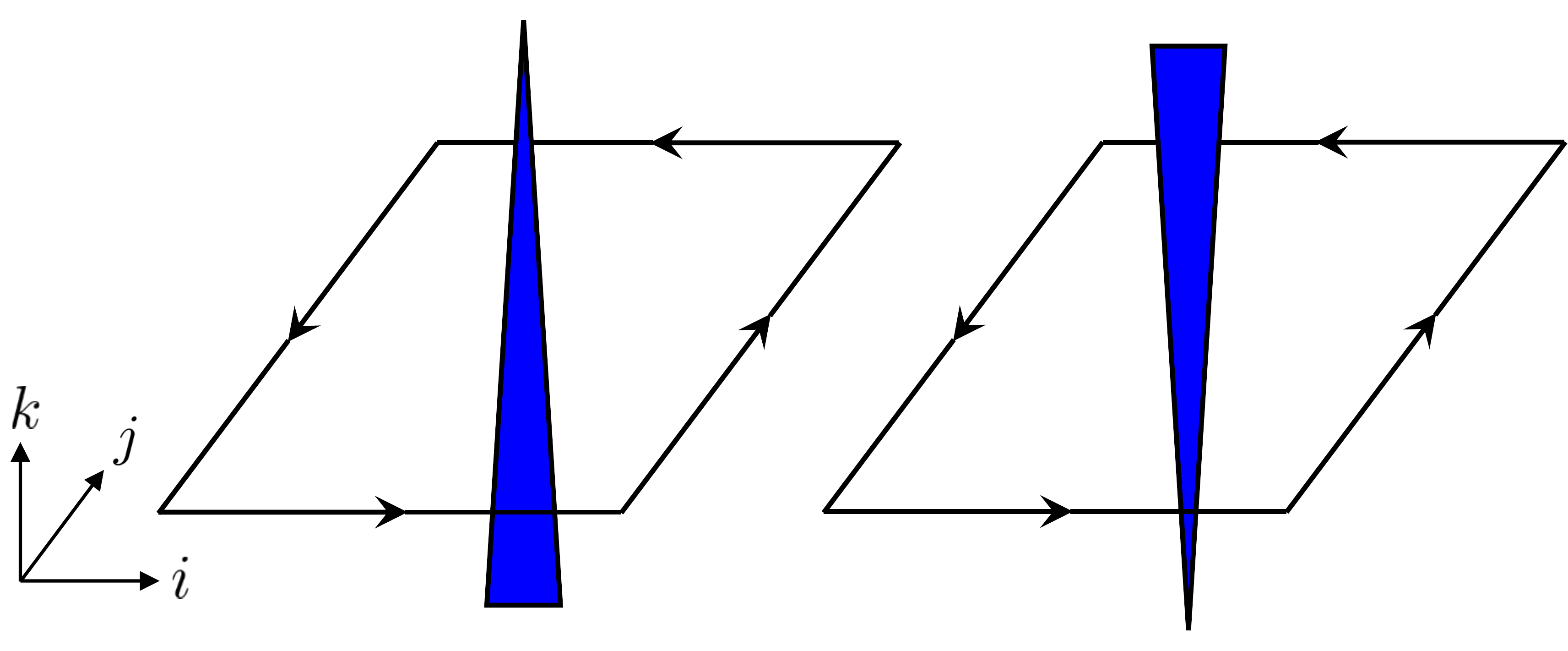}
	
	\vspace{-0.9em}
	
	\caption{\label{fig:visconvention} The visualisation convention for centre vortices. An \(m = +1\) vortex (\textbf{left}) is represented by a jet in the available orthogonal dimension, with the direction given by the right-hand rule. An \(m = -1\) vortex (\textbf{right}) is rendered by a jet in the opposite direction, such that we always show the flow of \(m = +1\) centre charge.}
\end{figure}

The visualisations are constructed following Ref.~\cite{Biddle:2019gke}. We slice through a dimension of the four-dimensional lattice by holding the selected coordinate fixed. Slices obtained by fixing temporal and spatial coordinates are referred to as ``temporal" and ``spatial" slices, respectively. The orientation of an \(m = +1\) vortex is determined by applying the right-hand rule, with an \(m = -1\) vortex rendered in the opposite direction. Since the flow of \(m = -1\) centre charge is indistinguishable from an opposite flow of \(m = +1\) centre charge, this implies the visualisations exclusively show the flow of \(m = +1\) centre charge. This convention is demonstrated in Fig.~\ref{fig:visconvention}.

\begin{figure*}
	\includegraphics[width=\linewidth]{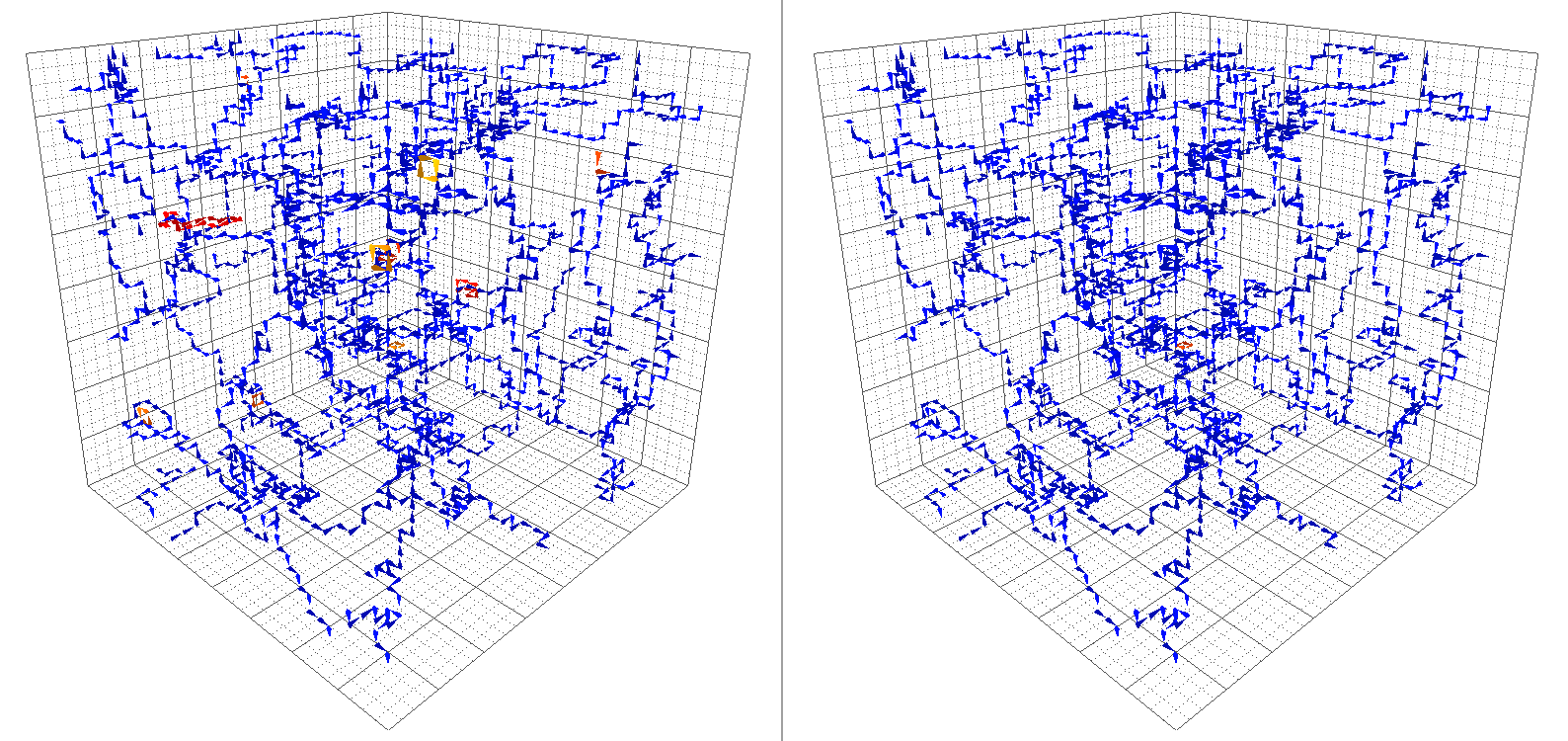}
	
	\vspace{0.8em}
	
	\includegraphics[width=\linewidth]{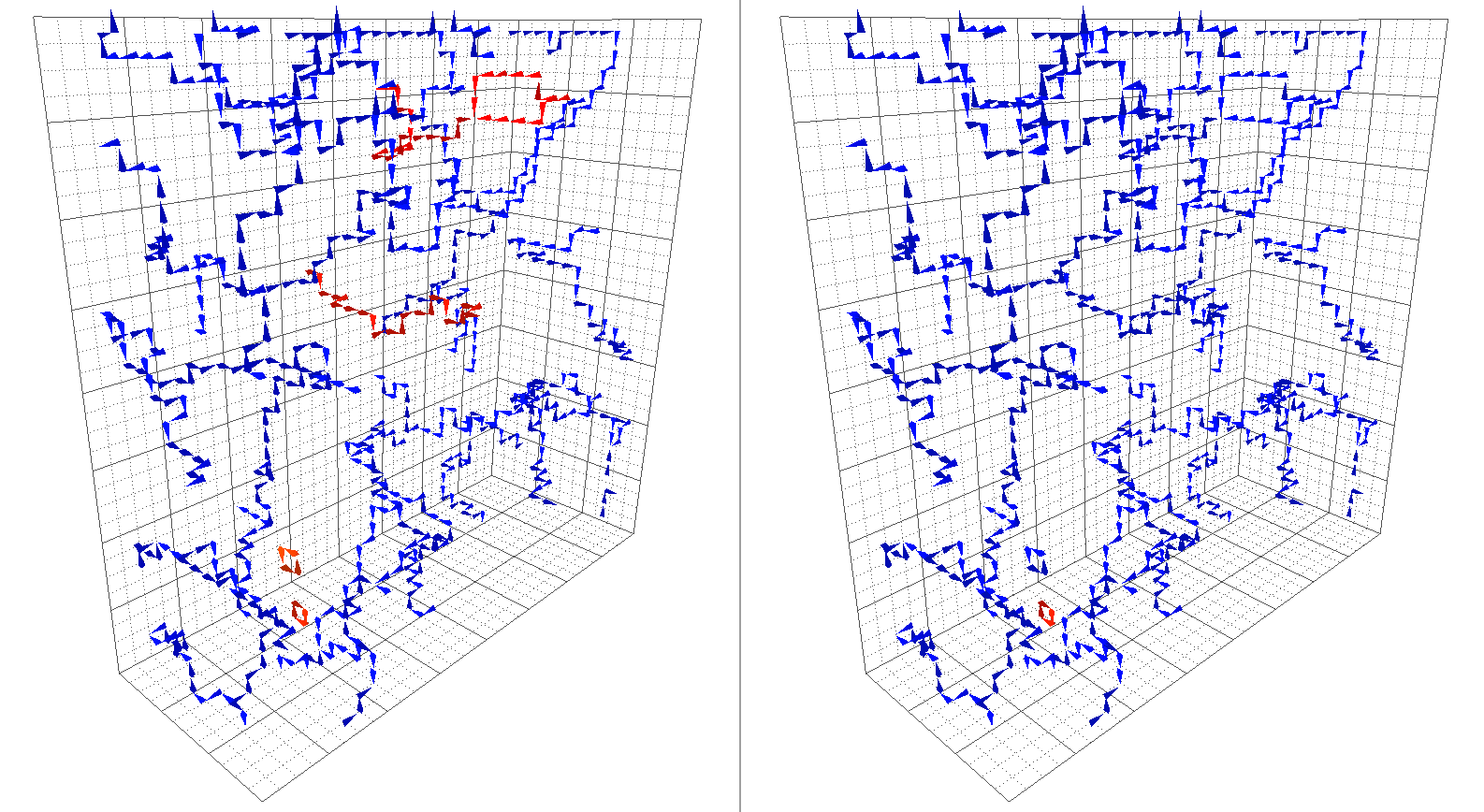}
	\caption{\label{fig:belowTdcomparison} A comparison of centre-vortex visualisations in the confined phase at \(T/T_d \simeq 0.55\) (\(N_\tau = 12\)) coloured by the loop in three dimensions (\textbf{left}) and sheet in four dimensions (\textbf{right}) to which each vortex belongs. An example temporal slice (\textbf{top}) and spatial slice (\textbf{bottom}) are shown. The majority of secondary loops are observed to lie in the same connected sheet in four dimensions. A single \(1 \times 1\) secondary loop is found to be disconnected from the percolating sheet in both cases.}
\end{figure*}

We start by comparing the two colour schemes in the confined phase, below \(T_d\). A side-by-side comparison of representative temporal and spatial slices is presented in Fig.~\ref{fig:belowTdcomparison}.
The left-hand side displays the usual setup, in which the percolating loop is coloured dark blue, and any separate secondary loops are shown with different colours. The right-hand side shows the new setup, with the vortex jets coloured by the sheet in which they lie. Immediately, we can notice that nearly all secondary loops, which initially appear disconnected in the three-dimensional slices, are connected in four dimensions. To be precise, they exist in the percolating \emph{sheet} that permeates all four spacetime dimensions below \(T_d\). This is reflected in the right-hand side of Fig.~\ref{fig:belowTdcomparison}, with the majority of jets that formerly had different colours now also coloured dark blue.

An example of vortices disconnected from the percolating sheet can be seen in both the temporal and spatial slices of Fig.~\ref{fig:belowTdcomparison}. In both cases this is a \(1\times 1\) vortex loop, the smallest possible such closed loop. They correspond to very small secondary sheets that exist in four dimensions. The number of these secondary sheets will be quantified in Sec.~\ref{subsec:secondarysheets}. For now, this points to the idea that below \(T_d\) any secondary loops disconnected from the percolating sheet tend to be very small, while the larger secondary loops are all connected in four dimensions. That said, it appears that even a majority of the \(1\times 1\) secondary loops are connected to the percolating sheet. This is especially apparent in the temporal slice shown in Fig.~\ref{fig:belowTdcomparison}. These would correspond to fluctuations in the sheet at the scale of the lattice spacing, small protrusions that result in an additional loop when slicing through the sheet. The proportion of such \(1\times 1\) loops that are connected and disconnected will be investigated in Sec.~\ref{subsec:loops}.

In Ref.~\cite{Biddle:2023lod} exploring the impact of dynamical fermions on centre-vortex structure, a proliferation of secondary loops was observed, including many small $1\times 1$ loops in the three-dimensional slices. The results presented here suggest that the main effect of dynamical fermions is to add additional fluctuations to the sheet structure in four dimensions. Thus, it will be important to examine this further in QCD.

\begin{figure*}
	\includegraphics[width=\linewidth]{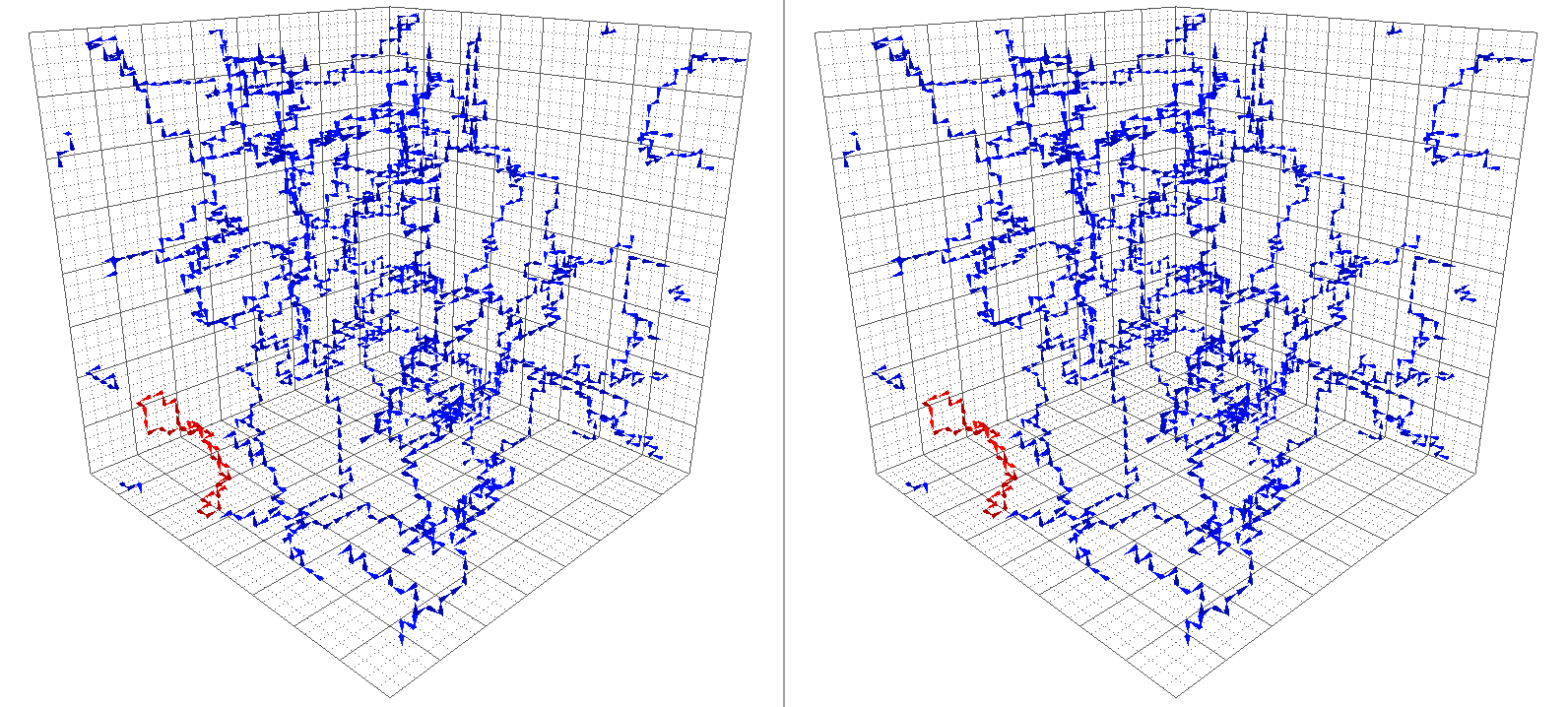}
	
	\vspace{0.8em}
	
	\includegraphics[width=\linewidth]{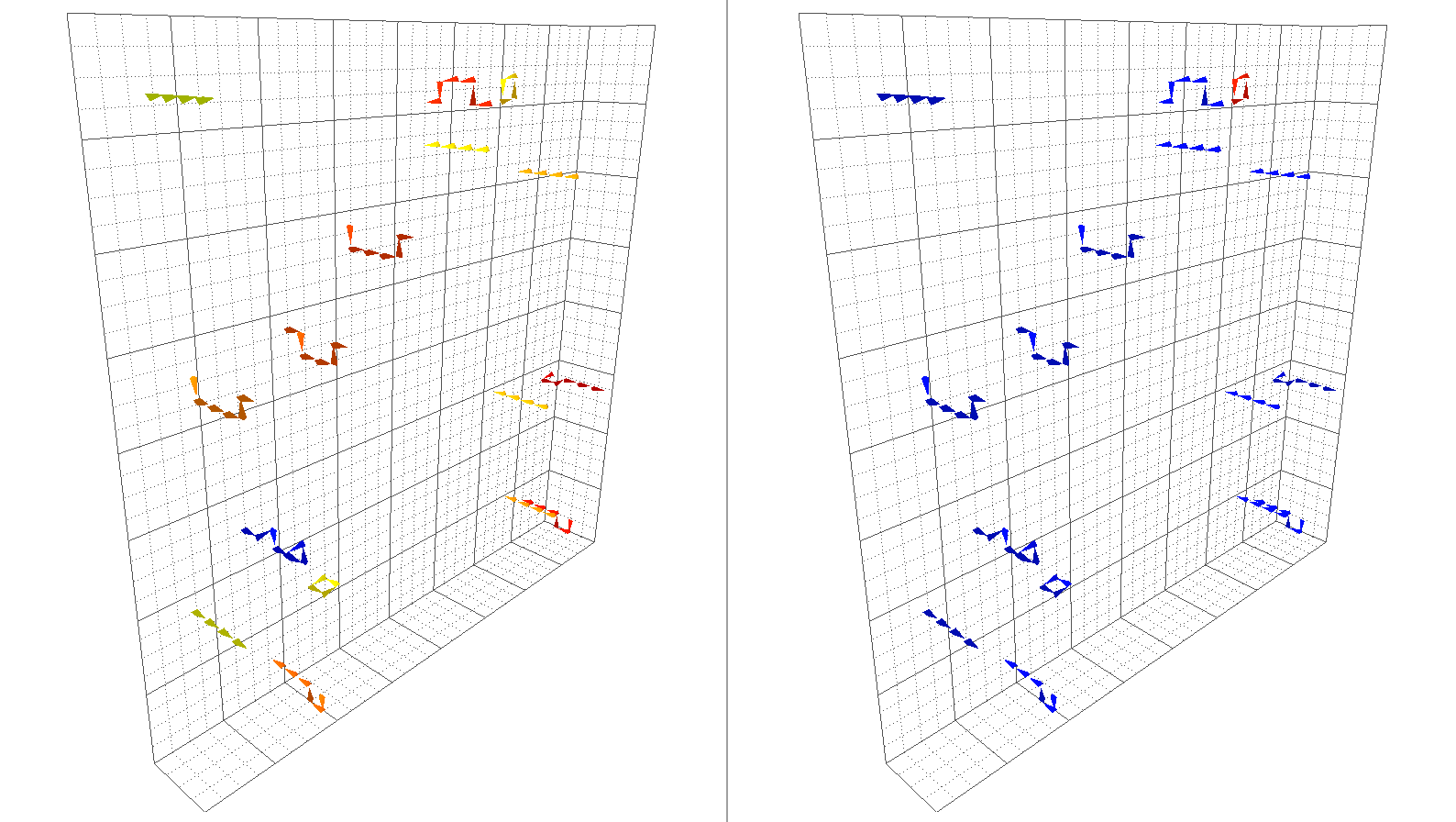}
	\caption{\label{fig:aboveTdcomparison} A comparison of centre-vortex visualisations in the deconfined phase at \(T/T_d \simeq 1.65\) (\(N_\tau = 4\)) coloured by the loop in three dimensions (\textbf{left}) and sheet in four dimensions (\textbf{right}) to which each vortex belongs. An example temporal slice (\textbf{top}) and spatial slice (\textbf{bottom}) are shown. Unlike below \(T_d\), a relatively large secondary loop is found to be disconnected from the percolating loop in the temporal slice. In spatial slices, the short vortex lines winding around the temporal dimension are primarily connected in four dimensions, embodying the fact that there is a single large sheet even at high temperatures.}
\end{figure*}

We now move into the deconfined phase, above \(T_d\). A similar comparison between temporal and spatial slices is presented in Fig.~\ref{fig:aboveTdcomparison}.
We recall that the principal change in centre-vortex geometry through the phase transition is an alignment with the temporal dimension. This is most easily seen in visualisations via spatial slices, comprising primarily short vortex lines winding around the temporal dimension. In the left-hand side of Fig.~\ref{fig:aboveTdcomparison}, these lines all have different colours. However, in the right-hand side they are all dark blue (with the exception of another \(1\times 1\) loop). This implies that even at very high temperatures, the centre-vortex structure is still dominated by a single large sheet. The lines only appear disconnected due to slicing through this sheet.

This could previously be inferred by additionally looking at temporal slices, in which a percolating loop is still formed. This indirectly implies that the vortex structure has not broken up into a large number of small sheets, but rather is still predominantly one connected cluster that has oriented along the temporal dimension. The updated\linebreak colour scheme provides an explicit representation of this fact, visually confirming that the percolating loop in temporal slices and the periodic lines in spatial slices are all part of the same structure.

For simplicity, we refer to the dominant sheet that exists at all temperatures as the ``primary sheet". It is interesting to consider whether the primary sheet is still percolating above \(T_d\) in spite of its temporal alignment. A simple test for percolation can be performed as follows. Define \(d_\mathrm{primary}\) to be the largest Euclidean distance between any two vortices (on the dual lattice) in the primary sheet. This is then normalised by the maximum \emph{possible} such distance \(d_\mathrm{max}\),
\begin{align}
	d_\mathrm{norm} &= \frac{d_\mathrm{primary}}{d_\mathrm{max}} \,, & d_\mathrm{max} &= \sqrt{\sum_{\mu = 1}^4 \left(\frac{N_\mu}{2}\right)^{\!\! 2}} \,,
\end{align}
where \(N_\mu\) is the lattice extent along dimension \(\mu\). If the\linebreak primary sheet has \(d_\mathrm{norm} = 1\) within a very small margin, then this signals that the sheet is percolating. This generalises a similar calculation that has previously been performed for loops in three-dimensional slices \cite{Engelhardt:1999fd, Biddle:2023lod, Mickley:2024vkm}.

We have calculated \(d_\mathrm{norm}\) at each temperature under consideration and found that it is exactly equal to 1 in every case. This verifies that, in the context of sheets in four dimensions, the centre-vortex structure is percolating at all temperatures.

Of course, due to the temporal alignment of the sheet, this percolation persists in three-dimensional slices above \(T_d\) only when slicing along the temporal dimension. In spatial slices, the percolating loop found below \(T_d\) gives way to lines of finite extent winding around the temporal dimension as \(T_d\) is crossed.

Staying with temporal slices, above \(T_d\) we find a comparatively large secondary loop that is genuinely disconnected from the primary sheet. This is in direct contrast to below \(T_d\), where larger loops were connected to the percolating sheet. We understand secondary loops that lie in the same sheet arise due to the sheet's curvature in four dimensions, as elucidated in Fig.~\ref{fig:disconnectedloops}. Above \(T_d\), with the vortex structure aligned with the temporal dimension, such instances are increasingly rare. This means one expects that any secondary loops are less likely to be connected at high temperatures, a fact that will be quantitatively verified in Sec.~\ref{subsec:loops}. That is, above \(T_d\) any larger secondary loops necessarily originate from disconnected sheets in four dimensions above.

\begin{figure}
	\centering
	\includegraphics[width=\linewidth]{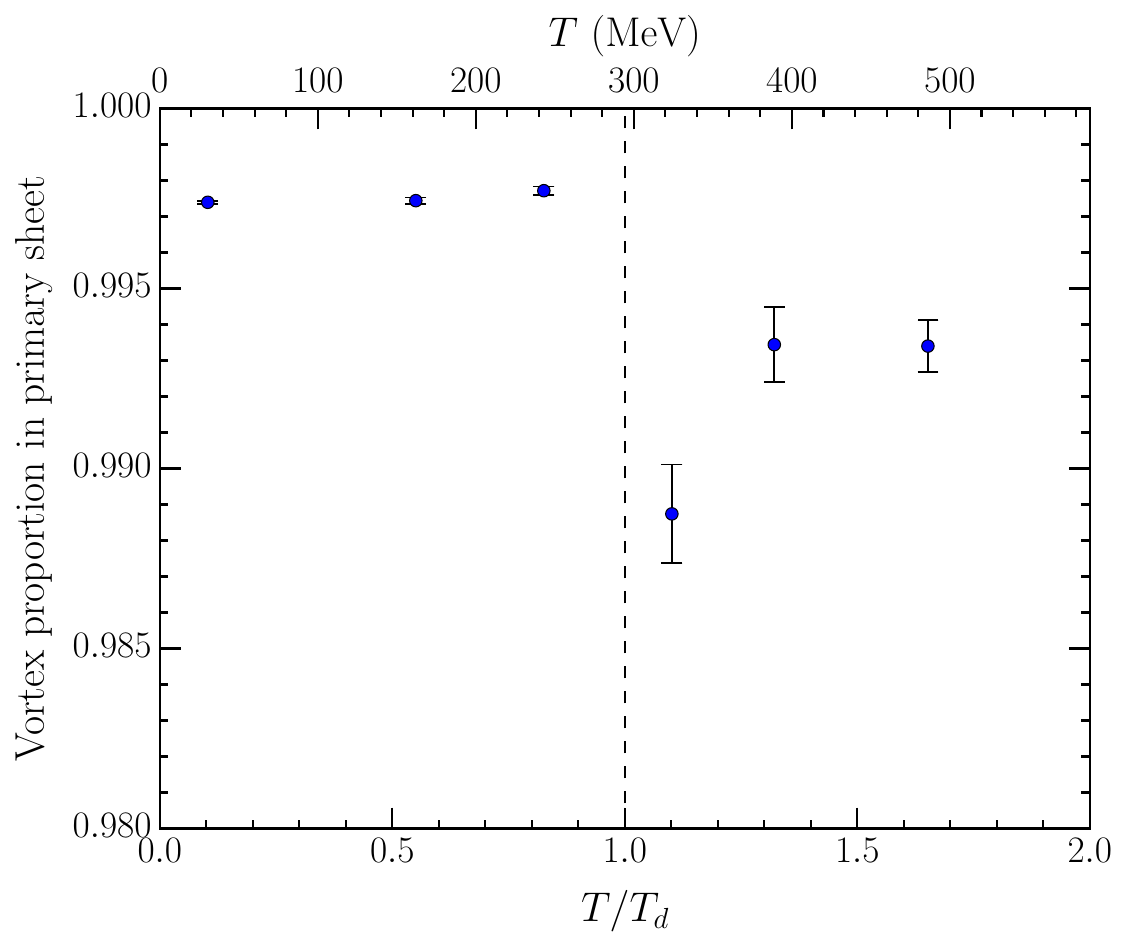}
	
	\vspace{-1em}
	
	\caption{\label{fig:primaryproportion} The proportion of vortices that belong to the largest (``primary") sheet. It sits consistently above 99\% throughout the full temperature range, barring a curious drop just above \(T_d\). The typically smaller proportion in the deconfined phase can be attributed to larger secondary sheets, as identified from the visualisations in Fig.~\ref{fig:aboveTdcomparison}.}
\end{figure}

Given that the centre-vortex structure comprises a primary sheet at all temperatures, an interesting quantity to examine is the proportion of vortices that belong to said sheet. The temperature evolution of this proportion is shown in Fig.~\ref{fig:primaryproportion}.
It generally attains values above 99\% regardless of temperature, reflecting the sheer density of the primary sheet. Still, a small statistically significant decrease is observed through \(T_d\). There are two ways this could manifest, either as an increase in the number of secondary sheets, or as an increase in their sizes. The high-temperature visualisations of Fig.~\ref{fig:aboveTdcomparison}, through which a large disconnected secondary loop was observed, suggest it is the latter for our highest temperatures.

The one outlier in Fig.~\ref{fig:primaryproportion} is our temperature nearest to but above \(T_d\), at which the proportion is notably smaller than all other temperatures. This point being an outlier is in fact a recurring feature throughout most quantities we have computed, and the reason for its existence will be explored in detail throughout Sec.~\ref{sec:analysis}.

\section{Vortex geometry analysis} \label{sec:analysis}
We now turn to a quantitative investigation of four-dimensional centre-vortex geometry, expanding on the qualitative aspects identified through visualisation. Having already quantified the relative size of the primary sheet that exists at all temperatures, in Sec.~\ref{subsec:secondarysheets} we will consider the number and sizes of secondary sheets. Subsequently, the four-dimensional connectedness of the loops seen in three-dimensional slices will be thoroughly examined in Sec.~\ref{subsec:loops}. All quantities are calculated using 100 configurations at each temperature, with statistical uncertainties obtained through bootstrapping.

\subsection{Secondary sheets} \label{subsec:secondarysheets}
\begin{figure}
	\includegraphics[width=\linewidth]{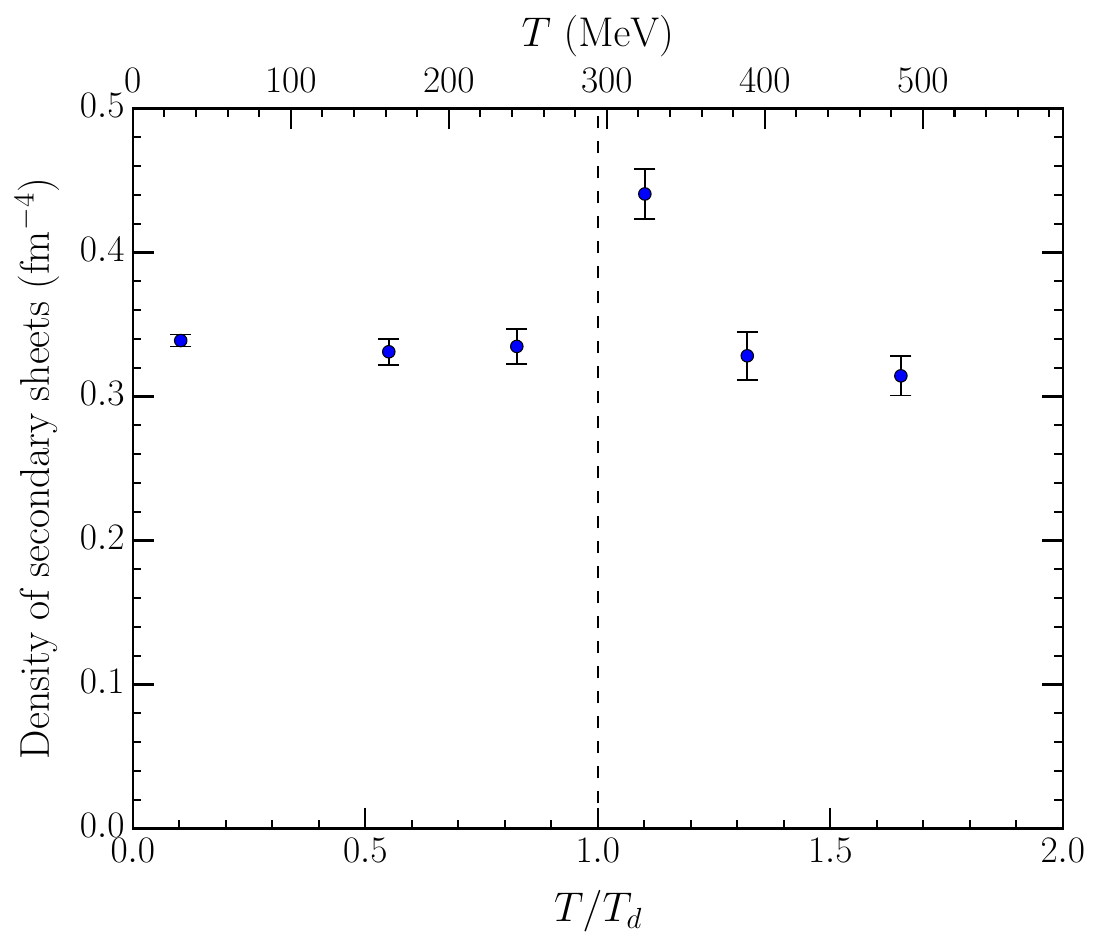}
	
	\vspace{-1em}
	
	\caption{\label{fig:secondarysheetdensity} The four-dimensional density of secondary sheets. It is approximately constant across the full temperature range, excepting a conspicuous increase just above \(T_d\). This is connected to the analogous decrease in proportion of vortices attributed to the primary sheet in Fig.~\ref{fig:primaryproportion}.}
\end{figure}

The first quantity we consider is the number of secondary centre-vortex sheets identified in four dimensions. This encompasses any sheet that is disconnected from the largest (primary) sheet. More precisely, to account for the reduction in four-dimensional volume as temperature increases, we compute a secondary-sheet density in fm\(^{-4}\) as the number of secondary sheets divided by the physical volume. This can then be compared across all temperatures. Its evolution is shown in Fig.~\ref{fig:secondarysheetdensity}.

We find that the four-dimensional density of secondary sheets is approximately constant. This again sets aside our temperature just above \(T_d\), at which there is a considerable spike in the density of secondary sheets. This spike naturally coincides with the decrease in proportion of vortices belonging to the primary sheet seen in Fig.~\ref{fig:primaryproportion}. With an increase in the number of secondary sheets relative to the volume, a greater proportion of vortices can be attributed to said sheets.

This could originate partly from the substantial reduction in total vortex density that is known to occur through the phase transition \cite{Mickley:2024zyg}. As less vortex matter fills the volume, more space is available for small secondary sheets to form. Since the vortex density is known to subsequently increase following this initial drop, this would primarily impact a small region just above \(T_d\). Another possible reason for the anomaly will be identified in Sec.~\ref{subsec:loops}.

It is noteworthy that the density of secondary sheets is otherwise constant (within statistical uncertainty) across the full temperature range. This value is rather small, only \(\simeq 1/3\)\thinspace fm\(^{-4}\). In particular, the density is approximately equal at both low and high temperatures. This is despite the small decrease in proportion of vortices in the primary sheet that persists at our highest temperature in Fig.~\ref{fig:primaryproportion}. This corroborates that the latter observation arises not from an increase in the \emph{number} of secondary sheets, but due to an increase in their typical \emph{size}.

\begin{figure*}
	\includegraphics[width=0.48\linewidth]{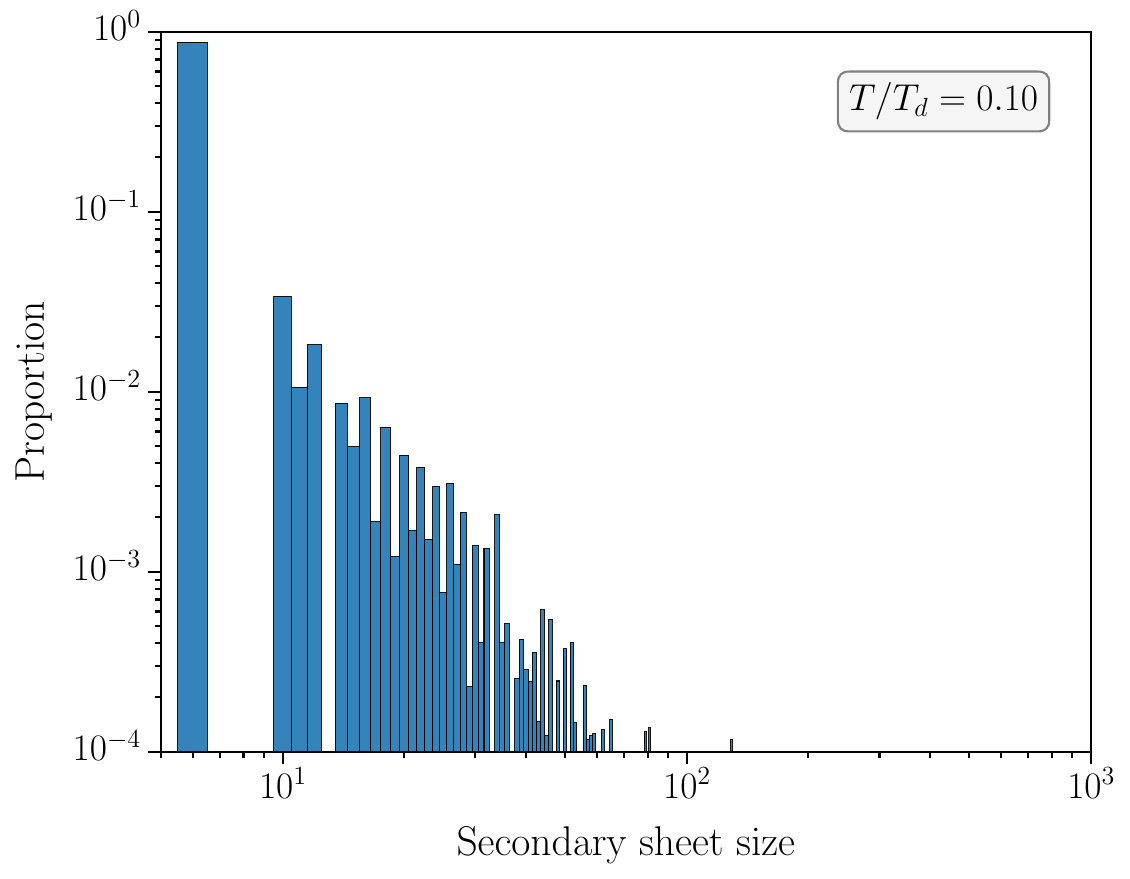}
	\hfill
	\includegraphics[width=0.48\linewidth]{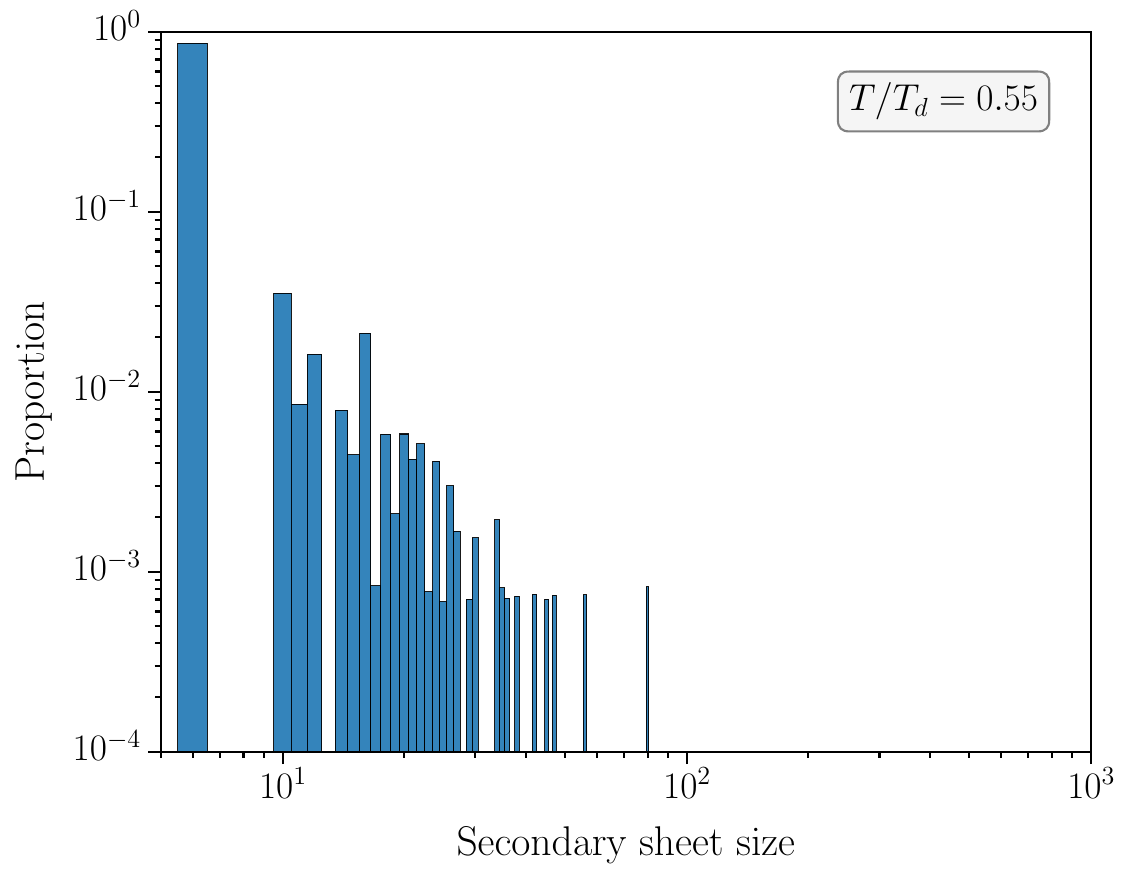}
	
	\vspace{1em}
	
	\includegraphics[width=0.48\linewidth]{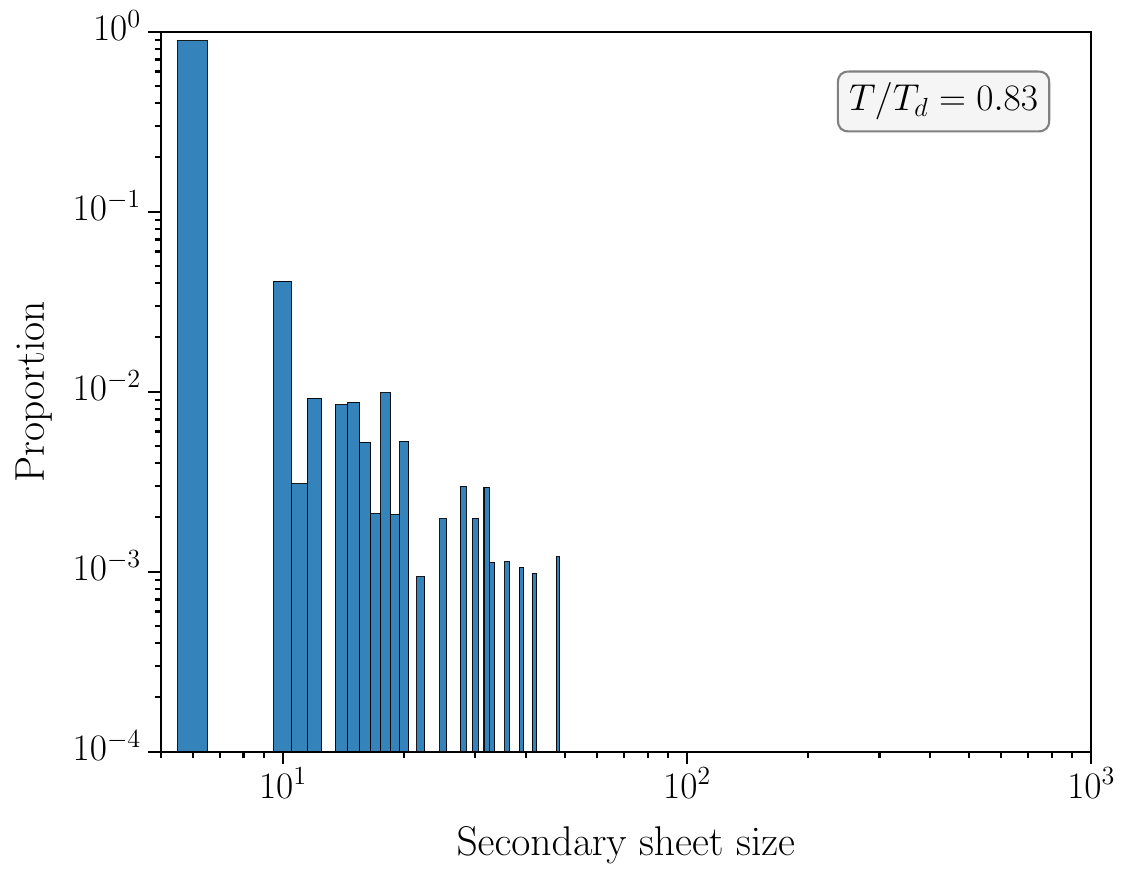}
	\hfill
	\includegraphics[width=0.48\linewidth]{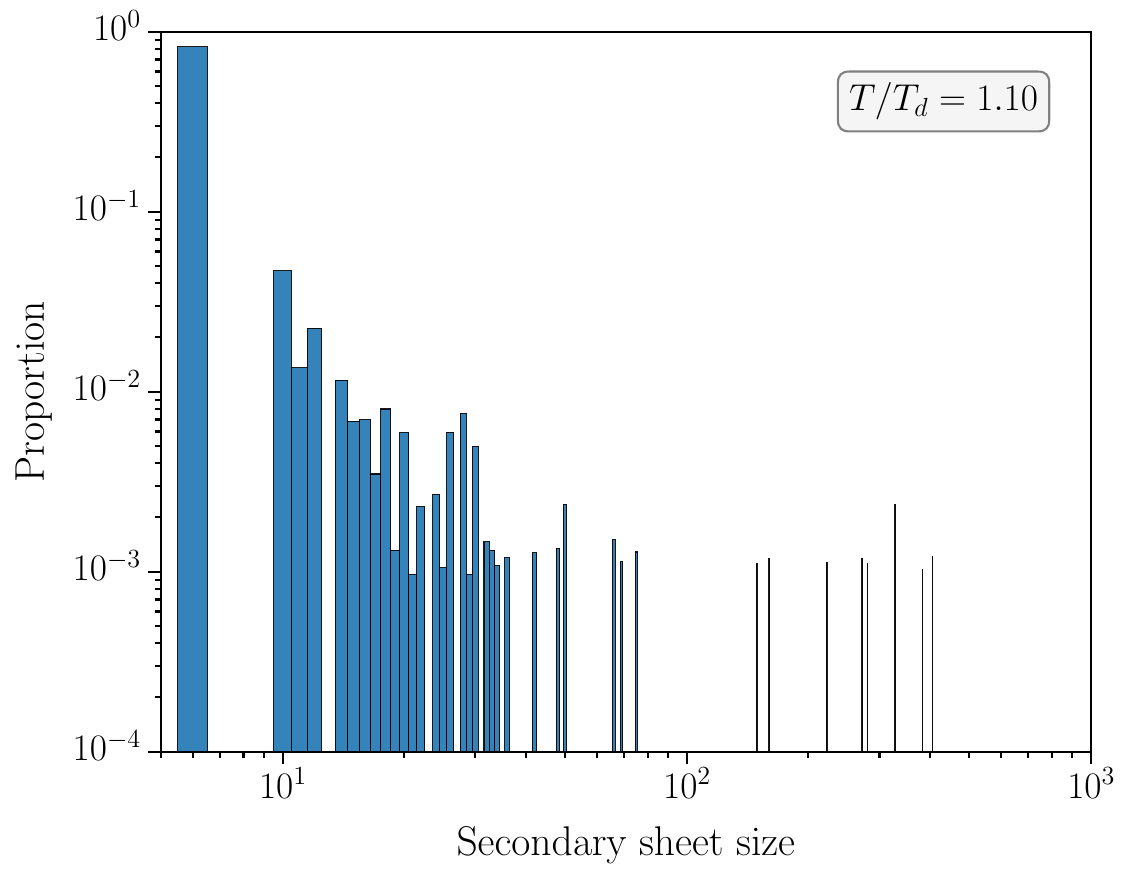}
	
	\vspace{1em}
	
	\includegraphics[width=0.48\linewidth]{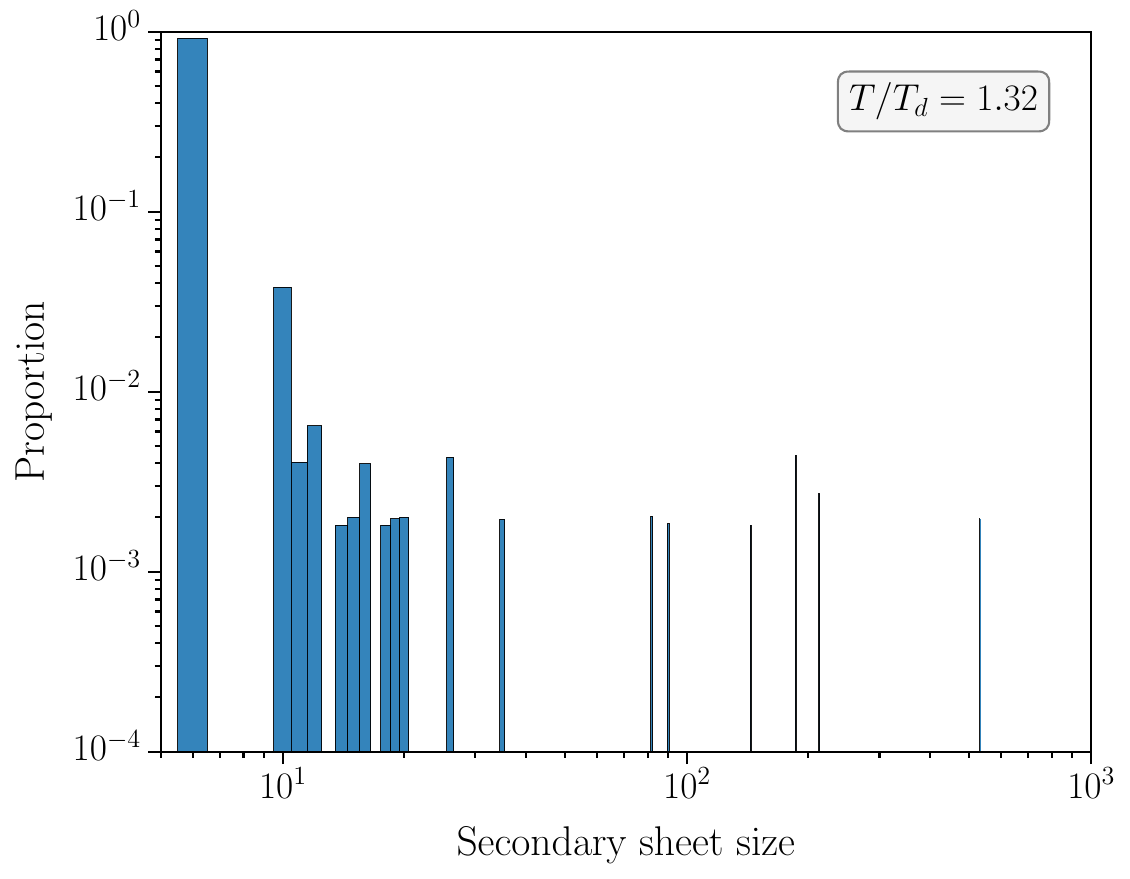}
	\hfill
	\includegraphics[width=0.48\linewidth]{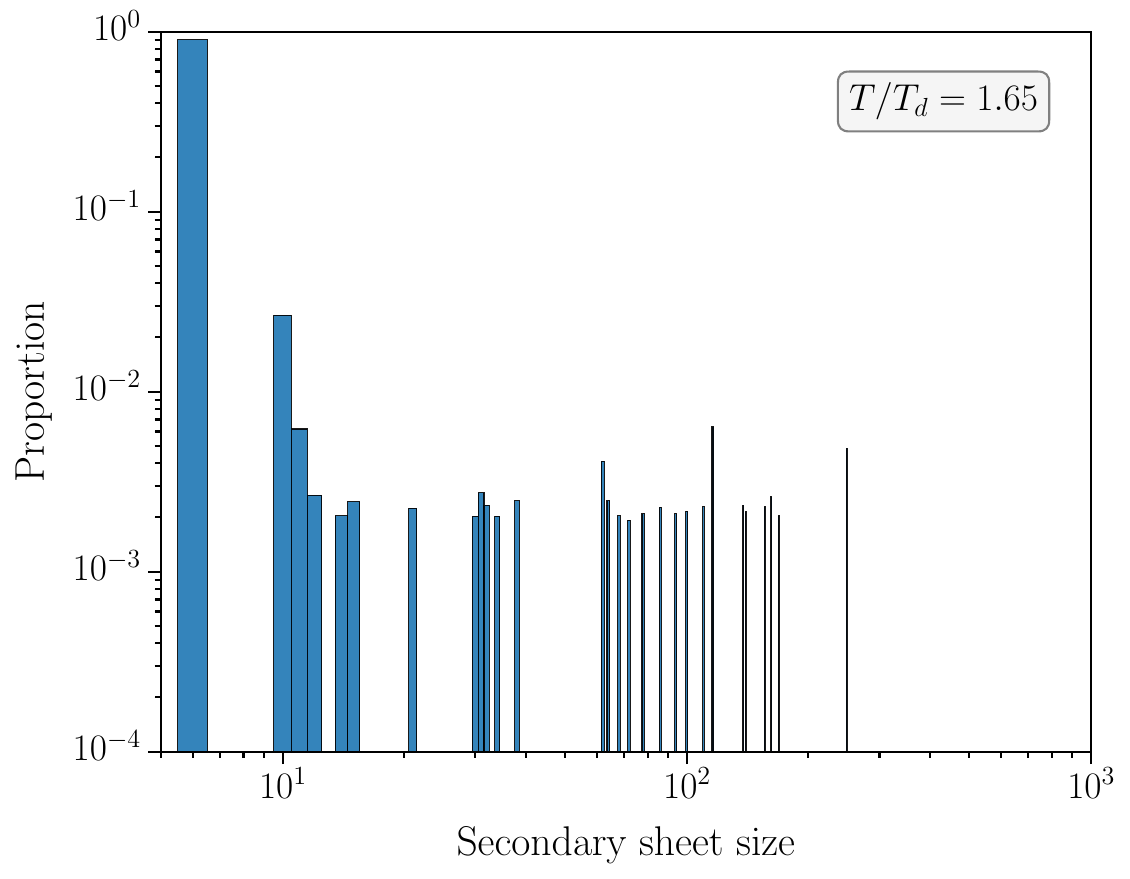}
	\caption{\label{fig:secondarysheetsizes} Histograms of secondary-sheet sizes on a log-log scale for \(T/T_d = 0.10\) (\(N_\tau = 64\), \textbf{upper left}), \(T/T_d = 0.55\) (\(N_\tau = 12\), \textbf{upper right}), \(T/T_d = 0.83\) (\(N_\tau = 8\), \textbf{middle left}), \(T/T_d = 1.10\) (\(N_\tau = 6\), \textbf{middle right}), \(T/T_d = 1.32\) (\(N_\tau = 5\), \textbf{lower left}), \(T/T_d = 1.65\) (\(N_\tau = 4\), \textbf{lower right}). The smallest and most common sheet size is six vortices, covering the six faces of a cube in the dual space. Sheets with an odd number of vortices are suppressed relative to their adjacent even sizes due to requiring at least one branching point somewhere in the sheet. Considerably larger secondary sheets are seen to appear above \(T_d\) compared to below \(T_d\).}
\end{figure*}

We now turn to examine these secondary-sheet sizes in detail. The number of vortices in each secondary sheet is counted, and histograms are produced to uncover the resulting distribution. These are shown for each temperature throughout Fig.~\ref{fig:secondarysheetsizes}.
Here, the vertical axis gives the proportion of secondary sheets with a given size. The histograms are given on a log-log scale. This allows the detail at small sheet sizes to be resolved, while simultaneously making visible the rare larger secondary sheet.

To start, the leftmost bin in each histogram is located at a size of six. This is the smallest number of vortices required to form a closed surface. It corresponds to a \(1\times 1\times 1\) (elementary) cube on the dual lattice. Such secondary sheets are far and away the most probable of any size, over an order of magnitude more prevalent than the next-smallest size. Slicing through these sheets produces the few disconnected \(1\times 1\) vortex loops found in the visualisations.

\begin{figure*}
	\raggedright
	\hspace{0.01\linewidth}
	\includegraphics[height=0.29\linewidth]{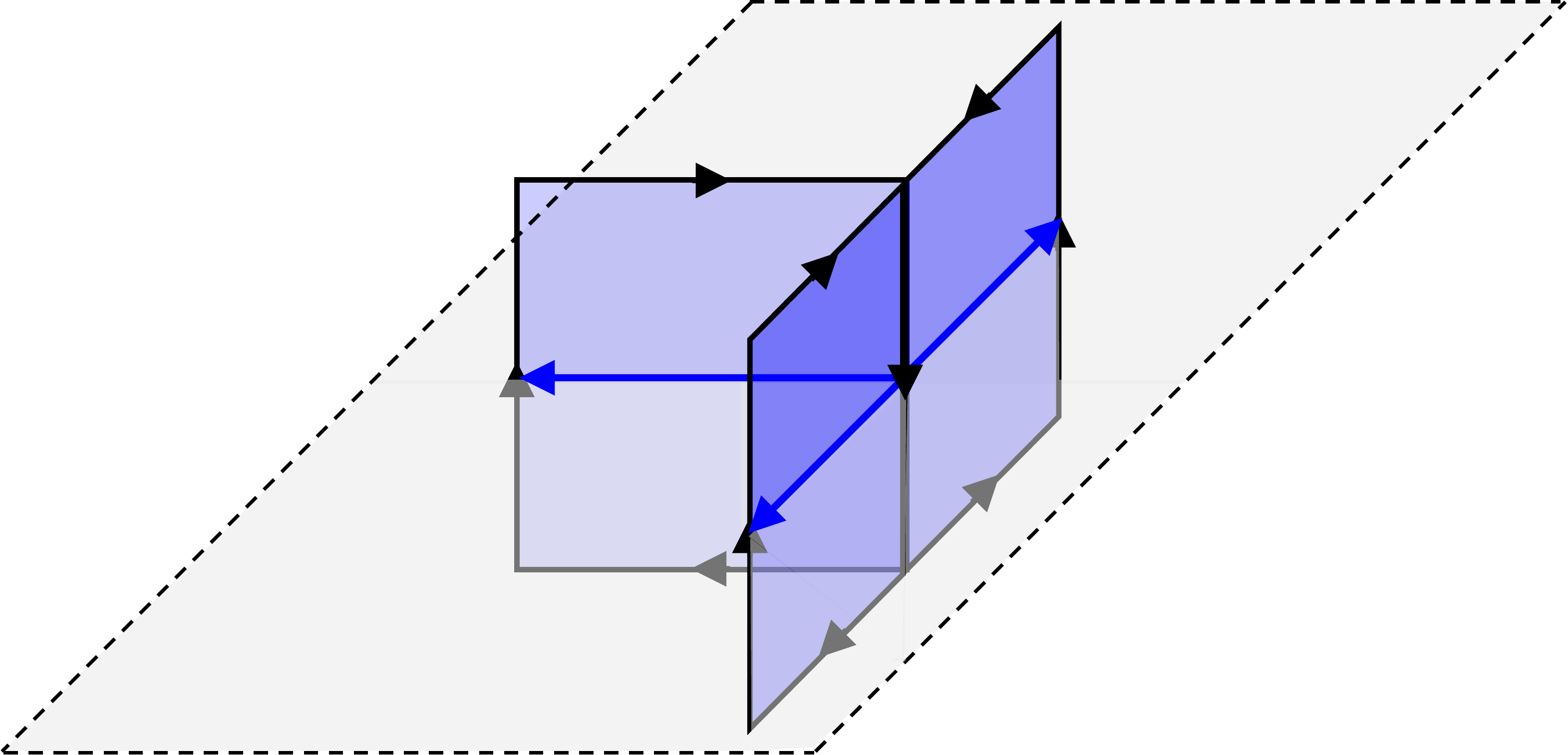}
	\hspace{0.08\linewidth}
	\includegraphics[height=0.29\linewidth]{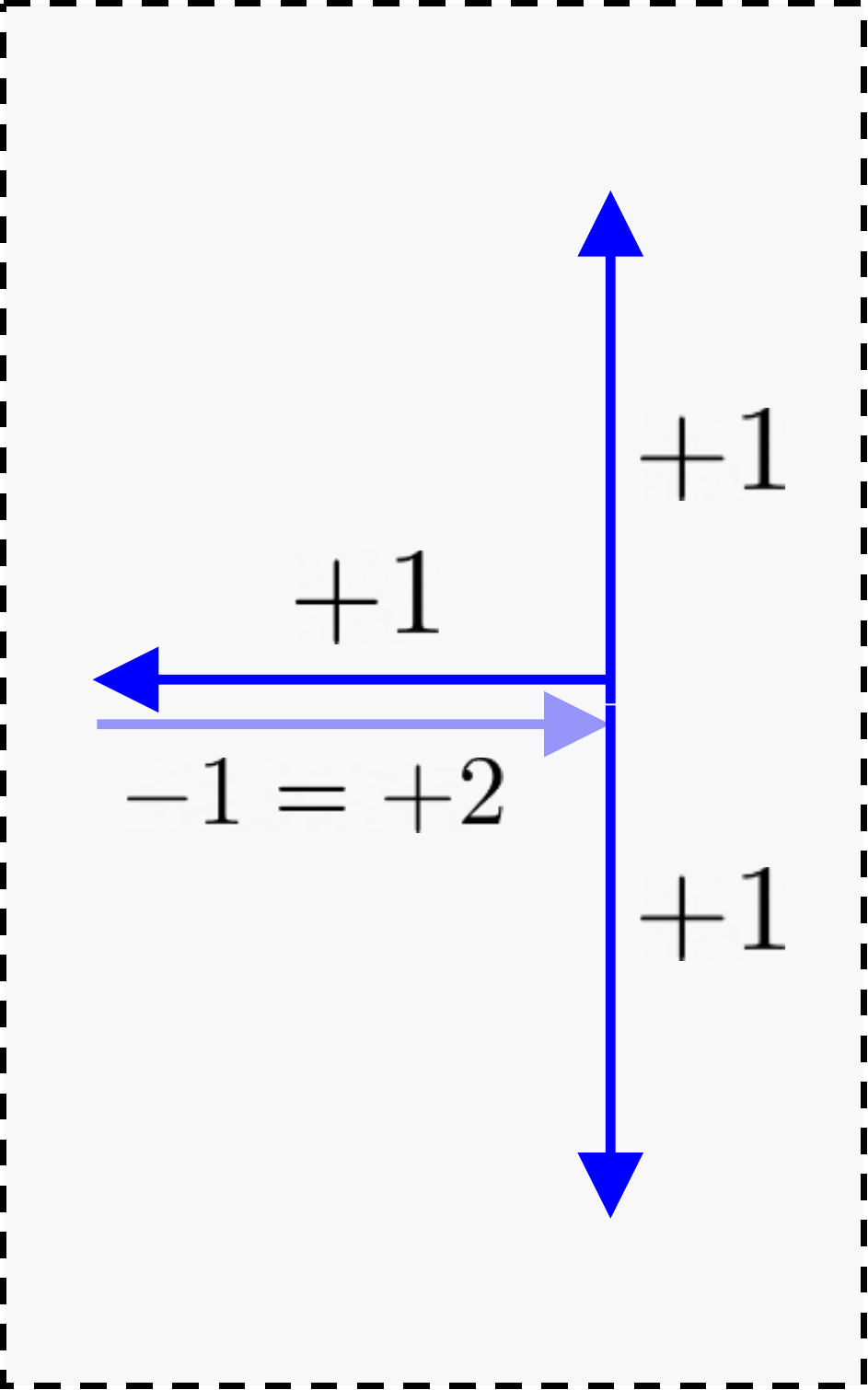}
	\caption{\label{fig:branchingpoint} Schematic of a vortex branching point/monopole. In four dimensions (\textbf{left}), monopoles occur when three faces of the vortex sheet (purple areas) connect at a common dual link. The black arrows around each face of the sheet are to indicate the vortex orientation, taken to be \(m = +1\). In taking a cross section through the sheet (i.e.\ ``slicing"), as portrayed by the dashed grey area, the monopole appears as three vortex lines emerging from a point (\textbf{right}). These arrows appear where the cross section intersects the vortex sheet. Examples of such monopoles can be found throughout the visualisations in Figs.~\ref{fig:belowTdcomparison} and \ref{fig:aboveTdcomparison}. Reversal of the left-hand arrow in the right figure instead shows the flow of \(m = -1\) centre charge, which is in turn equivalent to \(m = +2\) since \(-1 = +2\pmod 3\). Therefore, the right-hand diagram equivalently depicts the branching of centre charge.}
\end{figure*}
The next-smallest sheets comprise ten vortices. These primarily constitute (but are not limited to) \(1\times 1\times 2\) closed surfaces, and permutations thereof. Beyond a size of ten, there is an alternating trend where secondary sheets that comprise an even number of vortices are more probable than those with an odd number of vortices. This is because for a closed surface to contain an odd number of vortices, it necessarily features a branching point somewhere in the sheet. Centre-vortex branching points, also known as monopoles, are allowed because the centre charge is only conserved modulo \(N\) in general \(\mathrm{SU}(N)\) gauge theory. In \(\mathrm{SU}(3)\), this implies that three \(m = +1\) vortices can converge to or emerge from a point. The semantics of branching points are highlighted in Fig.~\ref{fig:branchingpoint}. They are relatively rare occurrences, such that sheets with an odd number of vortices are suppressed.

\begin{figure}
	\includegraphics[width=0.48\linewidth]{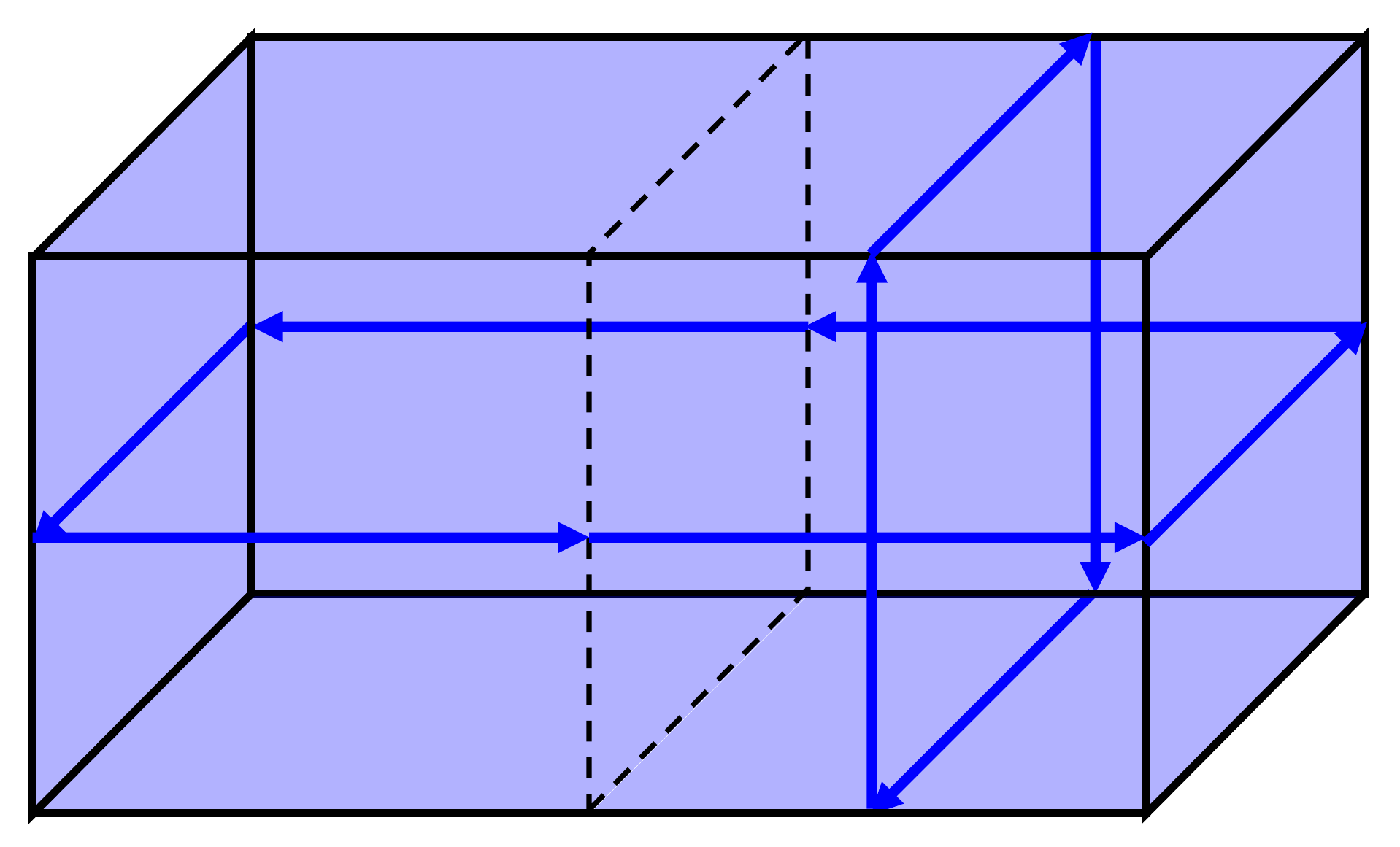}
	\hfill
	\includegraphics[width=0.48\linewidth]{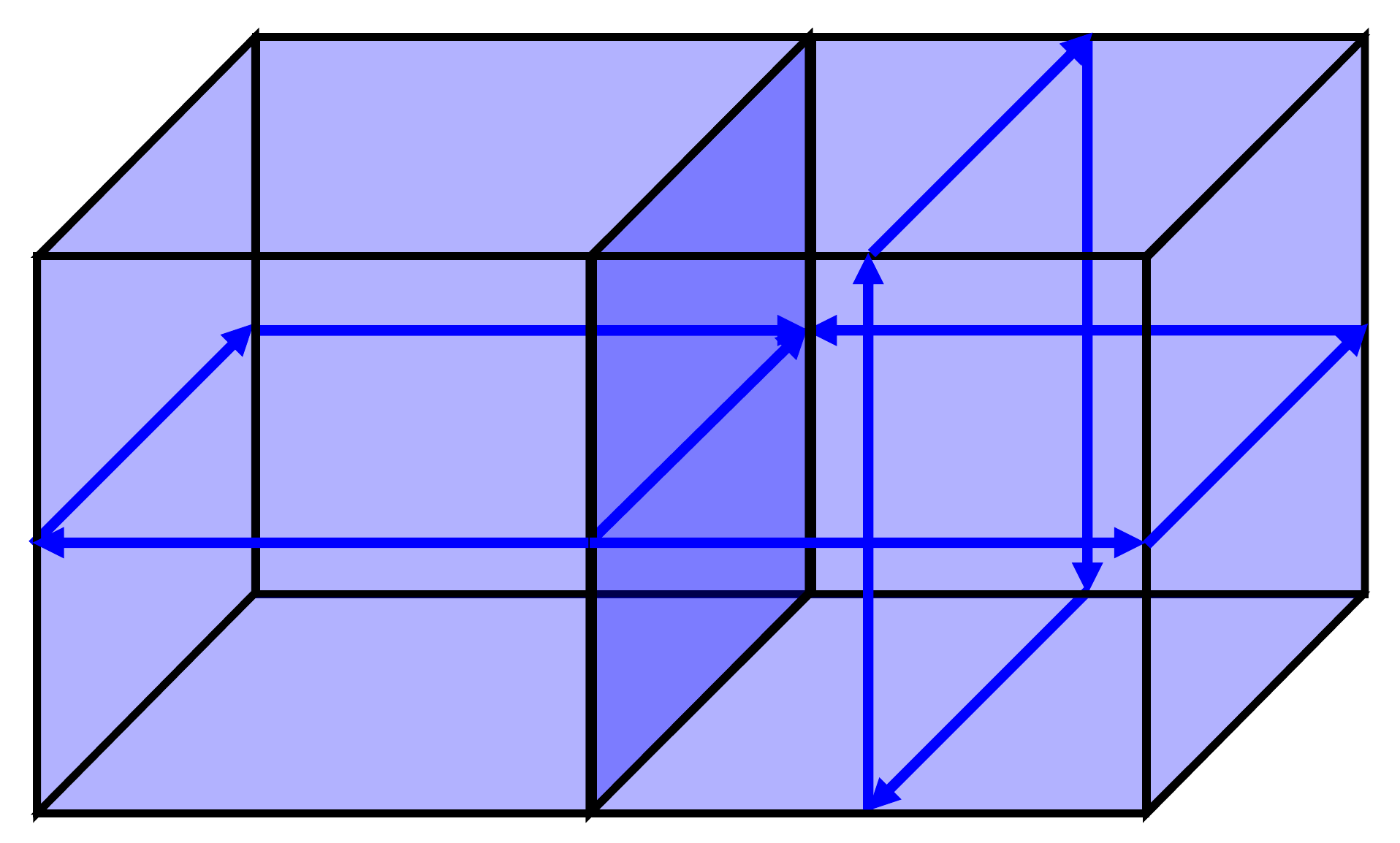}
	\caption{\label{fig:sizetenandeleven} Example vortex sheets of size ten (\textbf{left}) and eleven (\textbf{right}). The sheet of size ten is a \(2\times 1\times 1\) rectangular prism. In slicing through the sheet, one obtains either a \(2\times 1\) or \(1\times 1\) vortex loop depending on the dimension held fixed. The sheet of size eleven is also a rectangular prism, but with the middle face separating the two halves of the prism now part of the sheet and constituting a vortex in its own right. This therefore requires the existence of four branching points/monopoles (of the form illustrated in Fig.~\ref{fig:branchingpoint}), one for each edge of the middle face. Slicing horizontally through the sheet picks up two of these monopoles, in which it is clear that one monopole has three vortices emerging from a point, while the other has three vortices converging to a point.}
\end{figure}

As an example, sheets of size eleven are nearly identical to those of size ten, except the middle ``face" separating the two elementary cubes that form the \(1\times 1\times 2\) surface is also part of the sheet. This results in the formation of four branching points, one along each edge of the middle face, as demonstrated in Fig.~\ref{fig:sizetenandeleven}.

The primary shift in behaviour exhibited as the temperature climbs above \(T_d\) is the presence of larger sheets. Namely, the largest secondary sheet found below \(T_d\) is comprised of 129 vortices, while above \(T_d\) there are sheets with greater than several hundred vortices. This is despite the reduction in four-dimensional volume, providing explicit confirmation that appreciably larger secondary sheets are present in the deconfined phase. This accounts for the lower proportion of vortices in the primary sheet captured by Fig.~\ref{fig:primaryproportion}.

The appearance of larger secondary sheets might be attributed to the temporal alignment of the centre-vortex structure. While secondary sheets in the confined phase typically only exist for a couple of slices, above \(T_d\) they can easily be locked into winding around the temporal dimension and therefore would permeate a greater spacetime extent. This would also explain why the largest secondary sheet at our highest temperature (\(N_\tau = 4\)) is still smaller than that of the two lower temperatures (\(N_\tau = 5\) and \(6\)) with longer temporal extents.

It is also worth noting that, ignoring sheets with a size of six, the histogram at our lowest temperature appears approximately linear over even sheet sizes. Linearity on a log-log scale implies a power-law distribution for the sheet sizes,
\begin{equation} \label{eq:powerlaw}
	\log p(x) = \alpha - \beta \log x \implies p(x) = e^{\alpha} \, x^{-\beta} \,.
\end{equation}
This motivates performing fits of the form Eq.~(\ref{eq:powerlaw}) to the distribution of sheet sizes. Naturally, even and odd sizes cannot be simultaneously described owing to the suppression of the latter relative to the former. We exclusively consider even sizes due to their superior statistics.

\begin{figure}
	\includegraphics[width=\linewidth]{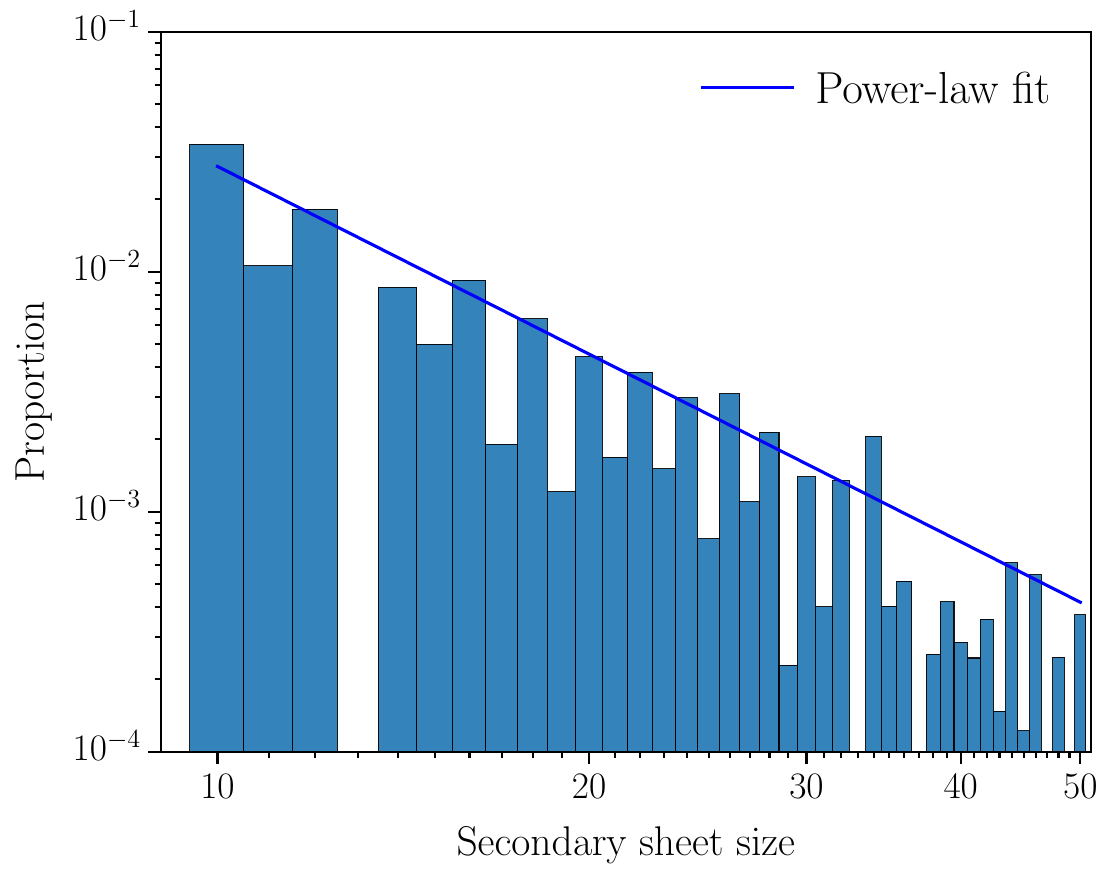}
	
	\vspace{-1em}
	
	\caption{\label{fig:powerlawfits} Power-law fit to the distribution of even secondary-sheet sizes at our lowest temperature \(T/T_d \simeq 0.10\) described in text, performed starting from a sheet size of twelve. The fit appears linear on the log-log scale. It is seen to adequately describe the data, particularly up to a size of \(\simeq 30\). Additional statistics are required to confirm whether a power law is a valid description of the secondary-sheet sizes.}
\end{figure}

In Fig.~\ref{fig:powerlawfits}, the lowest-temperature distribution of Fig.~\ref{fig:secondarysheetsizes} is reproduced over a more-limited range with our final fit overlaid. To suppress finite-volume effects, we consider secondary sheets up to a sheet size of 50.
We perform the fit starting from a minimum sheet size of twelve, which is found to better capture the behaviour at larger sizes. To ensure accurate parameter estimation, the fit is initially performed to the logarithm of the data, using the left half of Eq.~(\ref{eq:powerlaw}). The output is subsequently provided as initial guesses to a fit of the untransformed data, using the right half of Eq.~(\ref{eq:powerlaw}). This two-step procedure is essential, as taking the logarithm causes an implicit reweighting of the data in performing the fit. Thereby, this initially produces incorrect results that need to be corrected by the untransformed fit.

The fit is found to suitably describe the data within the displayed window, and is especially accurate up to sheet sizes of \(\simeq 30\) vortices. The power-law exponent is extracted as \(\beta \simeq 2.6\). It is interesting that we find \(\beta < 3\), given that a power-law distribution only possesses a finite variance for \(\beta > 3\). This is because of the very long tail exhibited by power-law distributions, which only converges to zero sufficiently fast to obtain a finite variance for \(\beta > 3\). Thus, if our value of \(\beta \simeq 2.6\) continues to describe the infinite-volume distribution, then we predict an infinite variance for the average secondary-sheet size.

Although these fits provide initial evidence that a power law can be used to describe secondary-sheet sizes, we note that additional statistics are required for verification. This is especially true if one wishes to perform these fits at higher temperatures, which induce less statistics by default due to the shorter temporal extent. Establishing whether empirical data follows a power law notoriously requires extreme statistics to capture the distribution's tail. This covers rare events at large sizes that nonetheless contribute significantly to statistical properties of the power-law distribution, such as its mean. Further inroads in this regard may be made with dynamical fermions, for which visualisations already reveal an abundance of secondary loops in three-dimensional slices compared to the quenched theory. It seems likely that this would partly be a consequence of an excess in secondary sheets, which may accordingly provide the statistics needed to accurately determine the distribution of sheet sizes.

\subsection{Connected and disconnected loops} \label{subsec:loops}
Having studied the vortex sheets that exist in four dimensions themselves, we now turn our attention to secondary loops in three-dimensional slices. Our goal is to understand the extent to which these secondary loops are connected in four dimensions (lie in the same sheet) and how this changes through the phase transition. Specifically, we are interested in the case where the loops lie in the primary sheet and hence form part of the dominant structure. With the primary sheet explicitly mapped out at all temperatures, this is a straightforward check. For simplicity, we will refer to any secondary loops that lie in the primary sheet as ``connected" and any that do not as ``disconnected".

Our focus for this analysis will be exclusively on temporal slices. The notion of ``secondary loops" is based on the existence of a percolating loop, which above \(T_d\) is only found in temporal slices (as seen in the visualisations of Fig.~\ref{fig:aboveTdcomparison}). This allows a direct comparison between the connectedness and disconnectedness of secondary loops above \(T_d\) to below \(T_d\), where a difference is expected due to the change in orientation of the vortex sheet.

\begin{figure}
	\includegraphics[width=\linewidth]{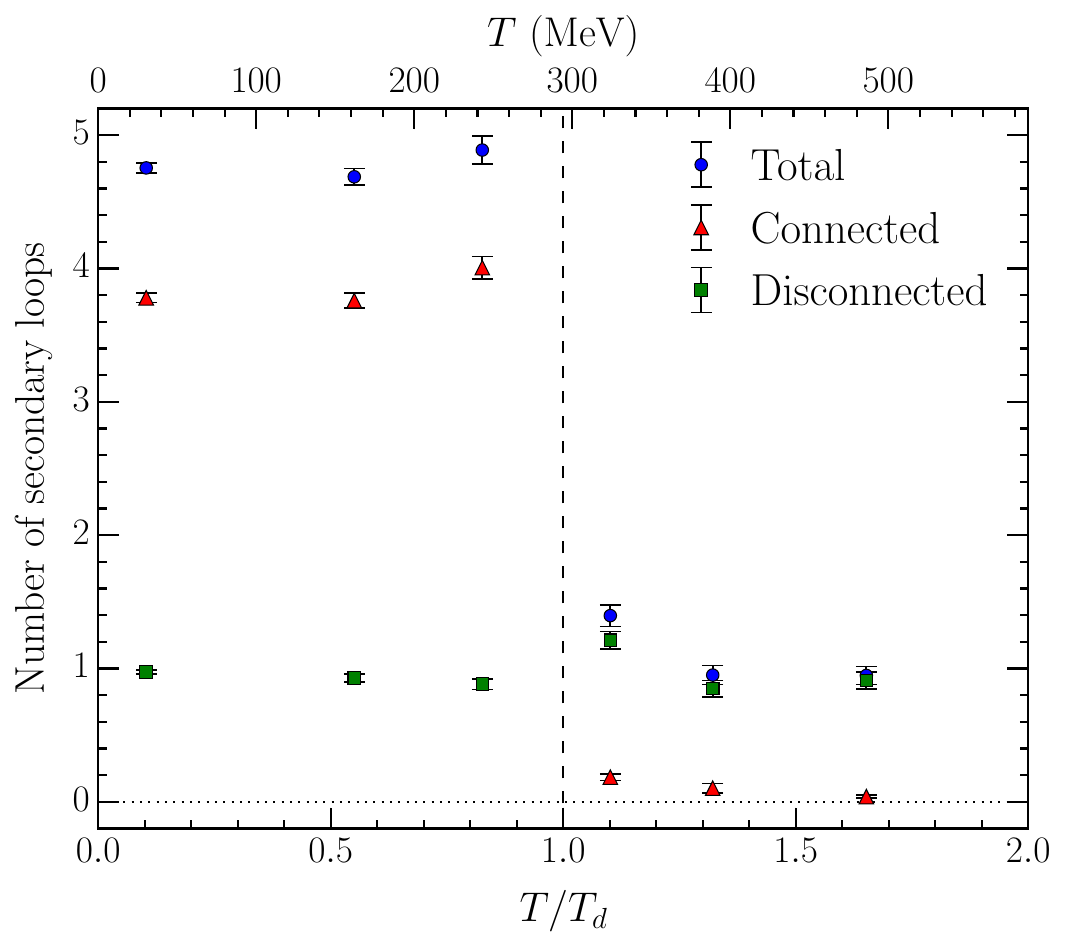}
	
	\vspace{-1em}
	
	\caption{\label{fig:secondaryloops} The average number of ``connected" (lie in the primary sheet) and ``disconnected" (lie in a secondary sheet) secondary loops per temporal slice. The combined average is also shown. Below \(T_d\), the majority of secondary loops are connected, arising from curvature of the primary sheet. Above \(T_d\), as the sheet aligns with the temporal dimension, the number of connected loops plummets to \(\simeq 0\). In contrast, the average number of disconnected loops is approximately constant across the full temperature range, with around \(\simeq 1\) disconnected loop per temporal slice.}
\end{figure}

To start with, the average number of connected and disconnected secondary loops per temporal slice is shown for each temperature in Fig.~\ref{fig:secondaryloops}.
The total number combining both connected and disconnected loops is also displayed for reference.

We find that below \(T_d\), the majority of secondary loops are connected to the primary sheet. This is unsurprising given the qualitative findings of Fig.~\ref{fig:belowTdcomparison}, in which comparing the two visualisation colour schemes revealed that most secondary loops are part of the percolating sheet. The number of disconnected secondary loops is comparatively small, averaging only \(\simeq 1\) such loop per temporal slice. This also matches the visualisations, in which a single \(1 \times 1\) disconnected loop can be observed.

\begin{figure*}
	\includegraphics[width=\linewidth]{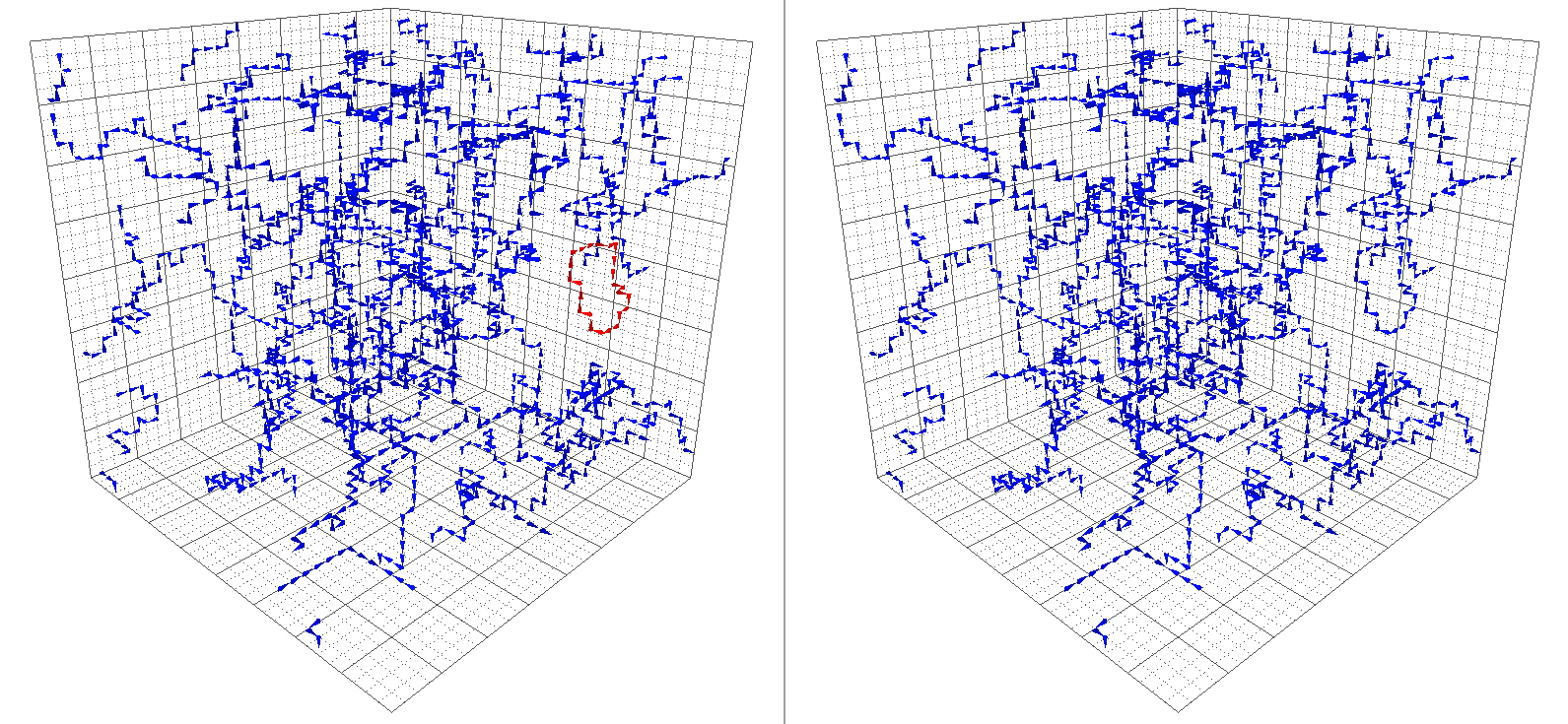}
	\caption{\label{fig:connectedloopaboveTd} An example of a rare connected secondary loop in temporal slices above \(T_d\) at \(T/T_d \simeq 1.65\) (\(N_\tau = 4\)), coloured by the loop in three dimensions (\textbf{left}) and sheet in four dimensions (\textbf{right}). Such occurrences are heavily suppressed above \(T_d\) due to the temporal alignment of the vortex sheet. This example persists only for the one slice, rejoining with the percolating loop in the following slice.}
\end{figure*}

As the phase transition is crossed, the average number of connected secondary loops rapidly falls to near zero, with such occurrences becoming very rare. This coincides with an equally steep drop in the combined average, as was initially observed in Ref.~\cite{Mickley:2024zyg}. We were able to find one instance of a connected secondary loop at our highest temperature, recorded in Fig.~\ref{fig:connectedloopaboveTd} for completeness.

This falloff has a natural explanation. As illuminated in Fig.~\ref{fig:disconnectedloops}, connected secondary loops appear because of curvature in the primary sheet. To be precise, for connected secondary loops to appear in temporal slices, the sheet must curve back on itself in the temporal dimension. However, above \(T_d\) the primary vortex sheet aligns with the temporal dimension. Therefore, such instances where the sheet curves to produce secondary loops in temporal slices become increasingly rare. This accounts for the reduction in secondary loops through \(T_d\). This was previously proposed as the reason behind this evolution in Ref.~\cite{Mickley:2024vkm}, though by identifying and counting the connected and disconnected loops we now have explicit confirmation.

In contrast, the typical number of disconnected secondary loop above \(T_d\) is roughly unchanged from below \(T_d\), still only \(\simeq 1\) per slice. As with previous quantities, this is excluding our temperature just above \(T_d\), which is slightly higher. In fact, the trend displayed by the number of disconnected secondary loops follows directly from the secondary-sheet density in Fig.~\ref{fig:secondarysheetdensity}. That is, if there is a greater density of secondary sheets in four dimensions, then there will on average be a greater number of disconnected secondary loops in three-dimensional slices. One can even derive an approximate relation between the two quantities. Denoting the secondary-sheet density by \(\rho_\mathrm{sheet}\), the number of secondary sheets is
\begin{align}
	N_\mathrm{sheet} = \rho_\mathrm{sheet} V \,, && V = N_s^3 N_\tau \, a^4 \,,
\end{align}
where \(V\) is the physical four-dimensional volume. Now, under the assumption that each secondary sheet spans only a single temporal unit (as in Fig.~\ref{fig:connectedloopaboveTd}), then it follows that the average number of disconnected secondary loops in temporal slices is
\begin{equation}
	N_\mathrm{disconnected} = N_\mathrm{sheet} / N_\tau = \rho_\mathrm{sheet} \, N_s^3 \, a^4 \,.
\end{equation}
Substituting in our approximate values of \(a \simeq 0.1\)\thinspace fm and \(\rho_\mathrm{sheet} \simeq 1/3\)\thinspace fm\(^{-4}\) gives \(N_\mathrm{disconnected} \simeq 1.1\). This is consistent with Fig.~\ref{fig:secondaryloops} to leading order. The true value is slightly less than 1.1, landing closer to exactly 1. This is likely due to sheets that exist at a single temporal coordinate on the dual lattice and therefore do not appear in any temporal slice, thus creating an overestimate. For instance, \(1\times 1\times 1\) elementary vortex sheets live within a three-dimensional subspace of the full four dimensions. If it is an \(x\)-\(y\)-\(z\) subspace, then this would produce a secondary loop in \(x\), \(y\) and \(z\) (i.e.\ spatial) slices of the lattice but not temporal slices.

Furthermore, Fig.~\ref{fig:secondaryloops} reveals additional context for the anomaly barely above \(T_d\). A similar increase is in fact also present below \(T_d\), but in the connected secondary loops. We also know from Ref.~\cite{Mickley:2024zyg} that there is a slight increase in vortex \emph{density} just below \(T_d\). Together, this makes it apparent that as \(T_d\) is approached from below, the extra vortex matter manifests as additional curvature in the primary sheet. This underlies the increase in connected secondary loops in this region below \(T_d\). Then, immediately as the phase transition is crossed and the alignment sets in, these additional regions of curvature effectively ``break off" from the primary sheet, generating a temporary spike in secondary sheets. This in turn realises an increase in the number of disconnected secondary loops found in temporal slices. The fact that this dynamic is only found in a small region around \(T_d\) suggests it would be interesting to investigate the finite-volume crossover in greater detail. We find it fascinating that our previous observations around \(T_d\) can be understood as curvature ``separating" from the primary sheet.

\begin{figure*}
	\includegraphics[width=0.48\linewidth]{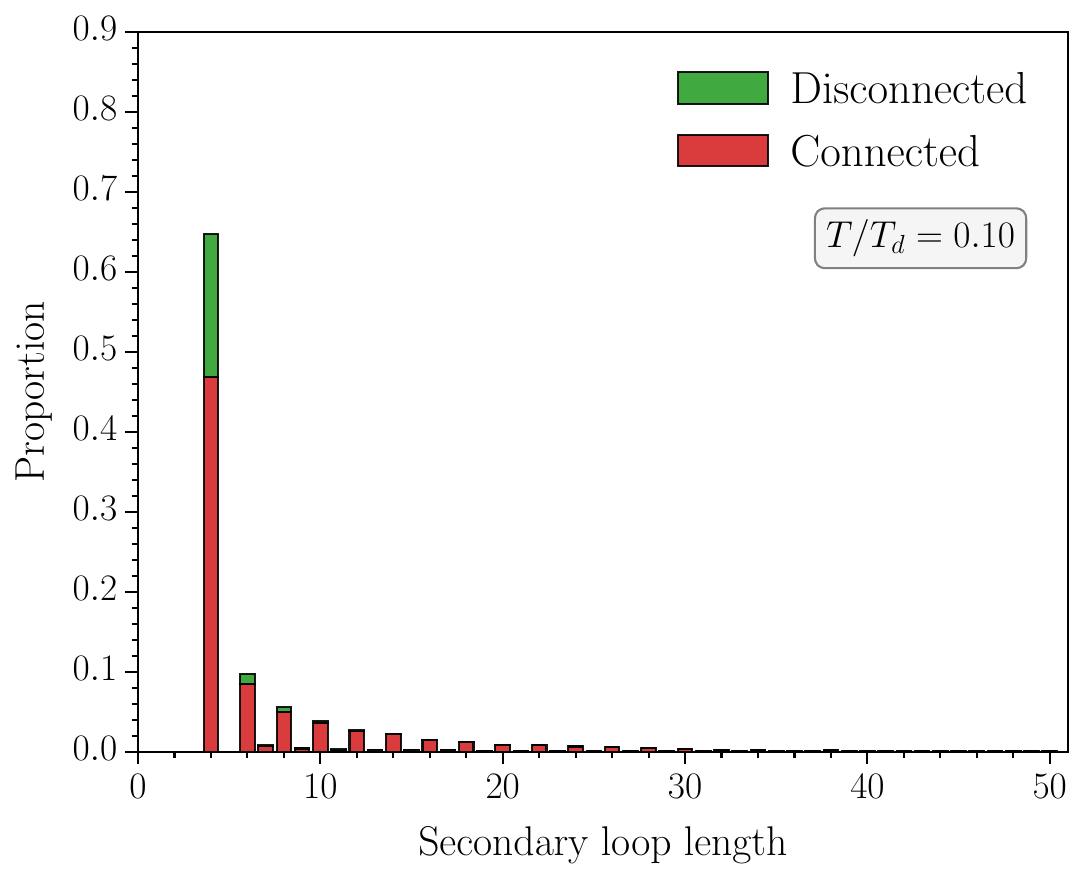}
	\hfill
	\includegraphics[width=0.48\linewidth]{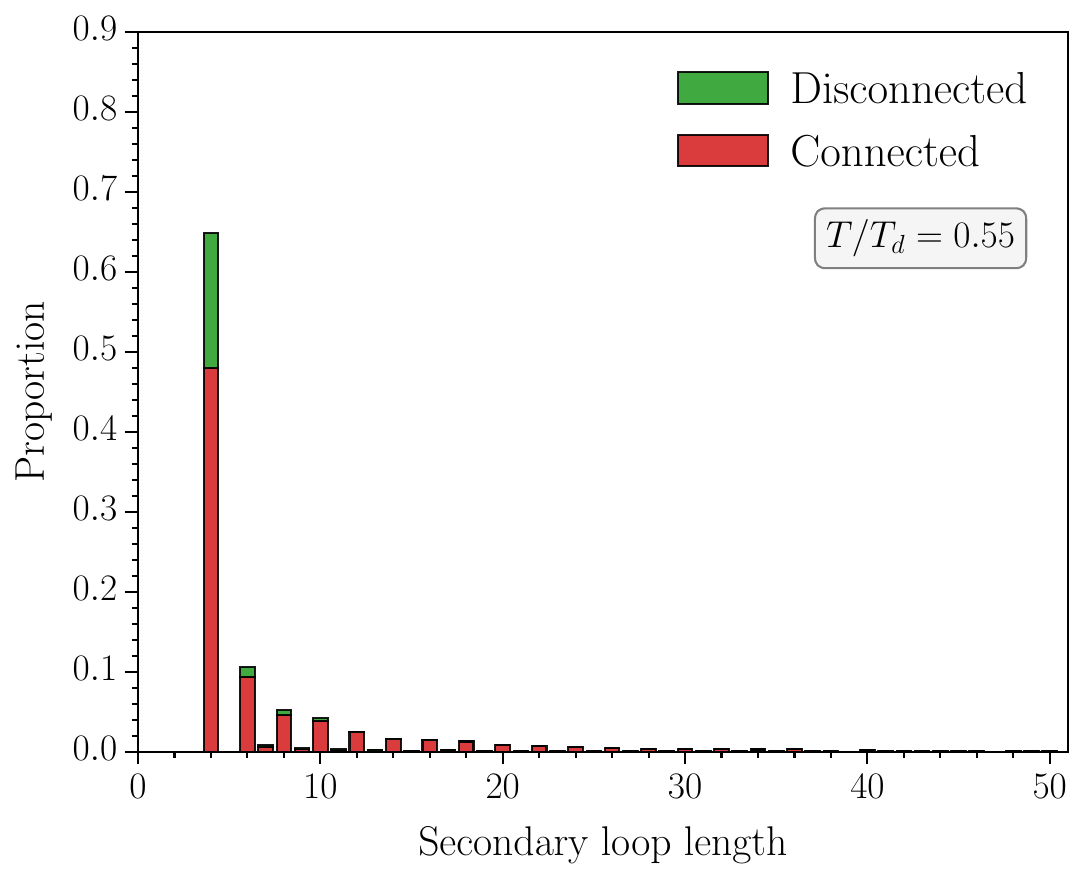}
	
	\vspace{1em}
	
	\includegraphics[width=0.48\linewidth]{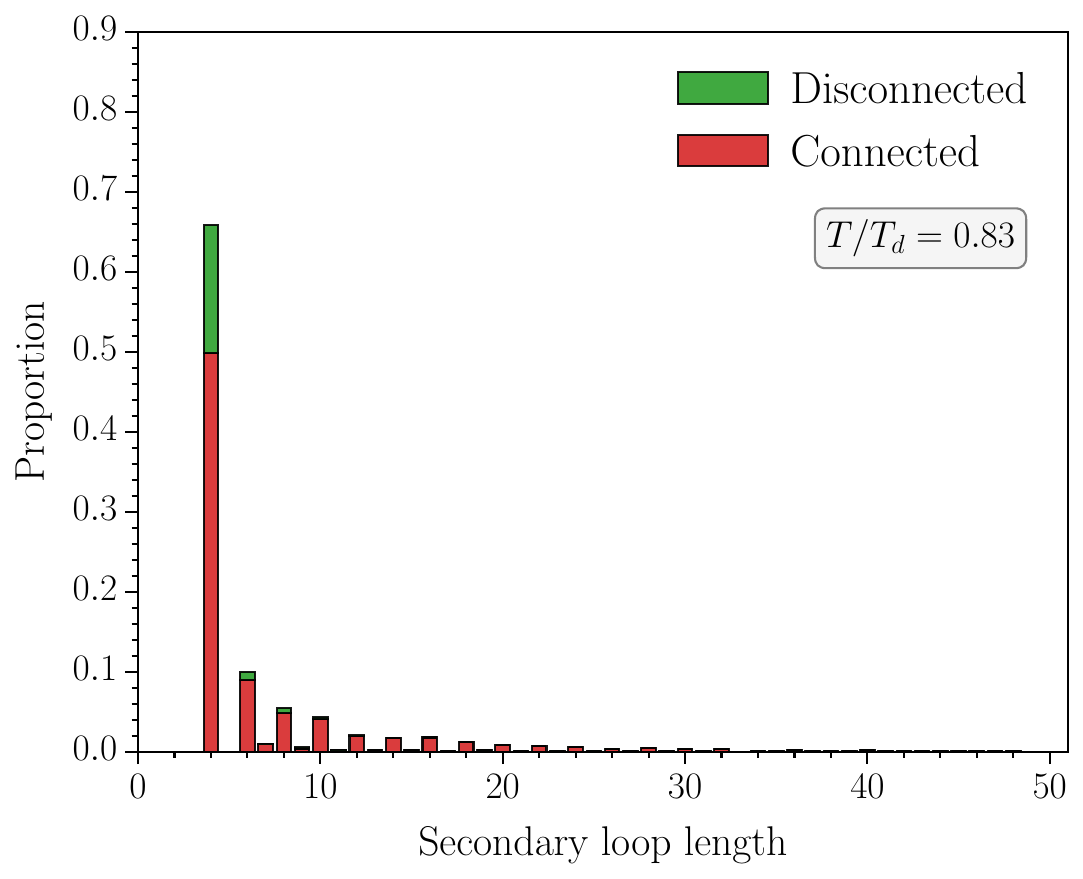}
	\hfill
	\includegraphics[width=0.48\linewidth]{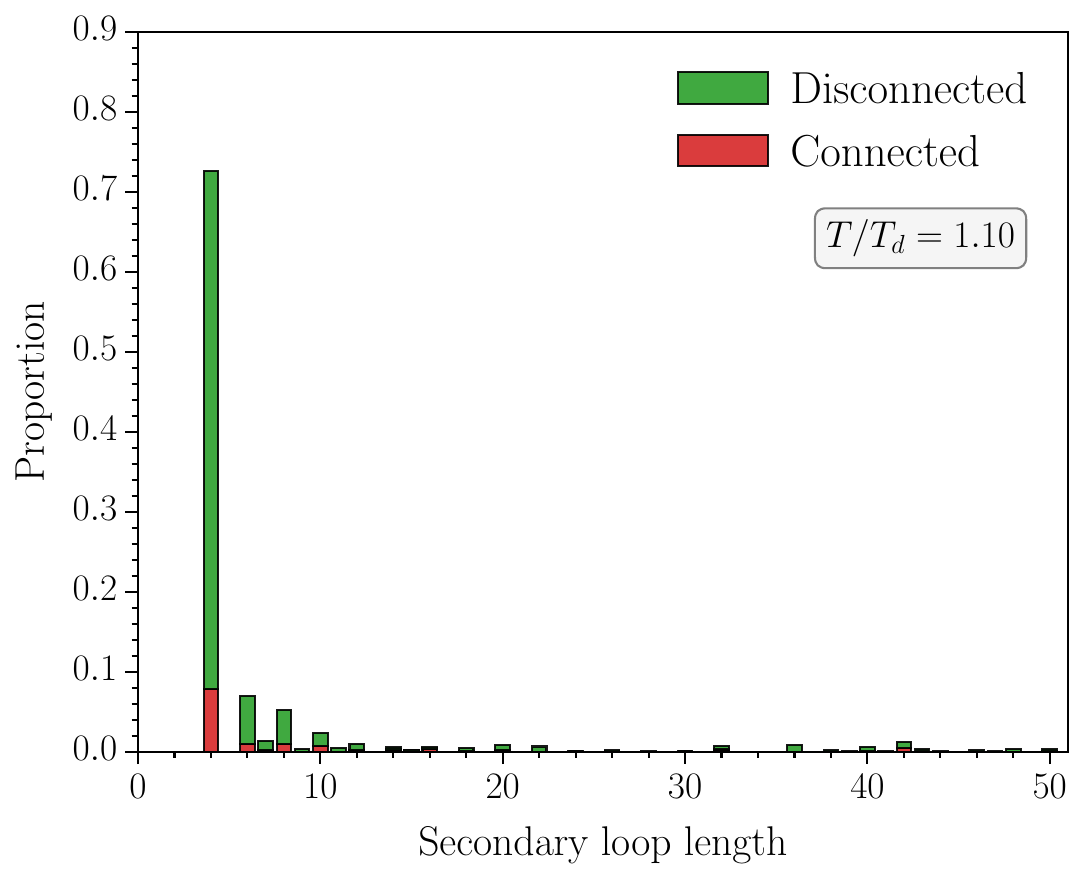}
	
	\vspace{1em}
	
	\includegraphics[width=0.48\linewidth]{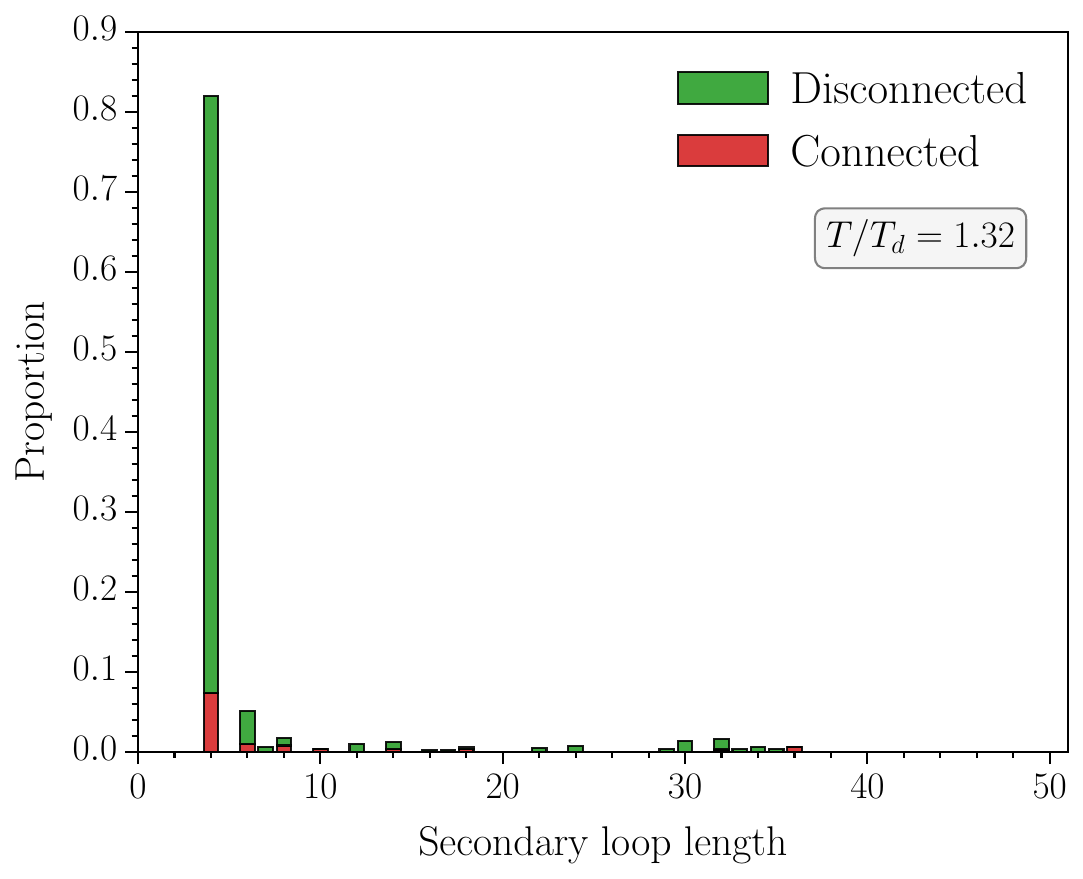}
	\hfill
	\includegraphics[width=0.48\linewidth]{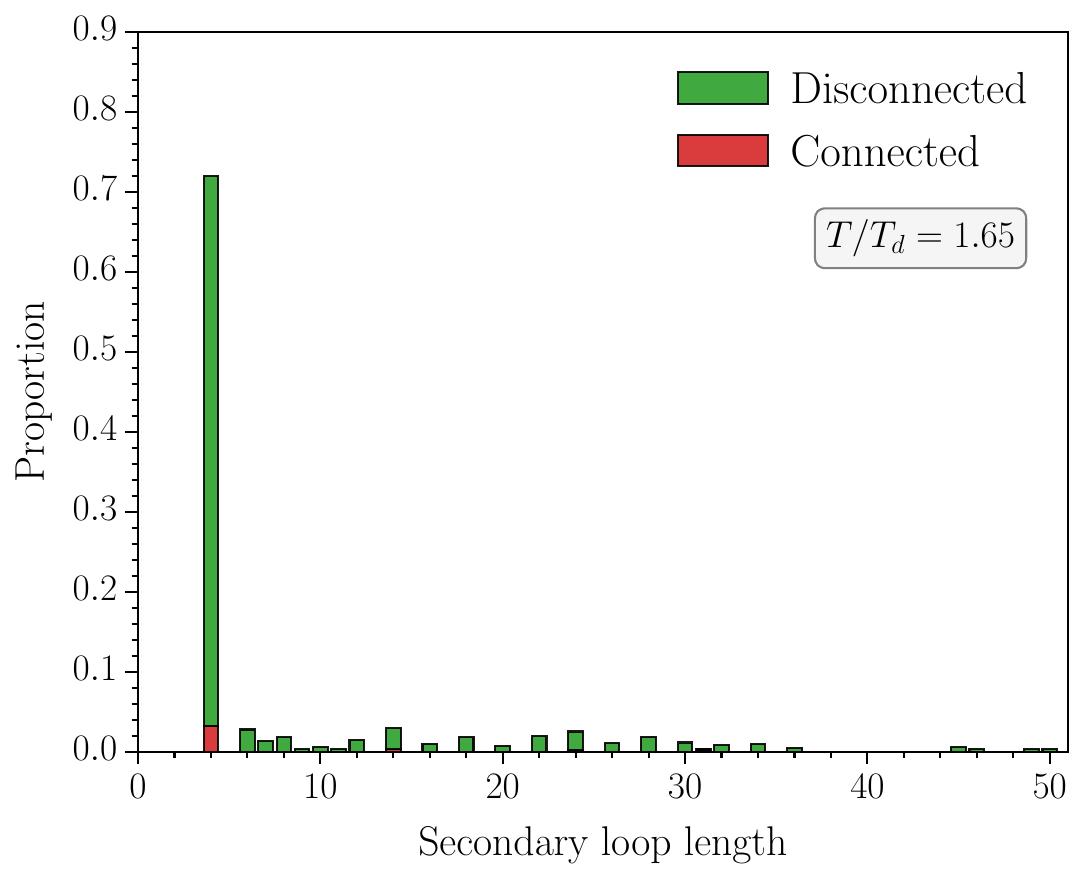}
	\caption{\label{fig:secondarylooplengths} Histograms of connected and disconnected secondary loop lengths for \(T/T_d = 0.10\) (\(N_\tau = 64\), \textbf{upper left}), \(T/T_d = 0.55\) (\(N_\tau = 12\), \textbf{upper right}), \(T/T_d = 0.83\) (\(N_\tau = 8\), \textbf{middle left}), \(T/T_d = 1.10\) (\(N_\tau = 6\), \textbf{middle right}), \(T/T_d = 1.32\) (\(N_\tau = 5\), \textbf{lower left}), \(T/T_d = 1.65\) (\(N_\tau = 4\), \textbf{lower right}). Below \(T_d\), a majority of secondary loops are connected, including elementary \(1\times 1\) loops of four vortices. The opposite is true above \(T_d\), where connected secondary loops of any length are suppressed.}
\end{figure*}

The final aspect of secondary loops we will consider is their lengths, again restricted to temporal slices. The number of vortices in each connected and disconnected secondary loop is counted and a histogram is made in a similar vein to the sheet sizes in Fig.~\ref{fig:secondarysheetsizes}. Instead of showing separate histograms for connected and disconnected loops, a single stacked histogram is produced. This allows the relative proportions of connected and disconnected loops of a given length to be easily discerned. These histograms are presented in Fig.~\ref{fig:secondarylooplengths}.

We find that the histograms below \(T_d\) are dominated by connected secondary loops. This is expected for larger lengths, but it is interesting that even the elementary \(1\times 1\) secondary loops that comprise four vortices are more likely to be connected than disconnected. The example temporal slice visualised in Fig.~\ref{fig:belowTdcomparison} hints at this, though it was unclear whether it would hold true generally. We find an approximate \(3:1\) split of \(\mathrm{connected}:\mathrm{disconnected}\) elementary vortex loops below \(T_d\).

With connected secondary loops suppressed above \(T_d\), the histograms are instead dominated by disconnected secondary loops at all lengths. The occasional connected \(1\times 1\) loop is found at a rate of about 1 in 10 just above \(T_d\), and down to about 1 in 20 at our highest temperature. Any connected secondary loops longer than four vortices are vanishingly rare.

\begin{figure*}
	\includegraphics[width=0.48\linewidth]{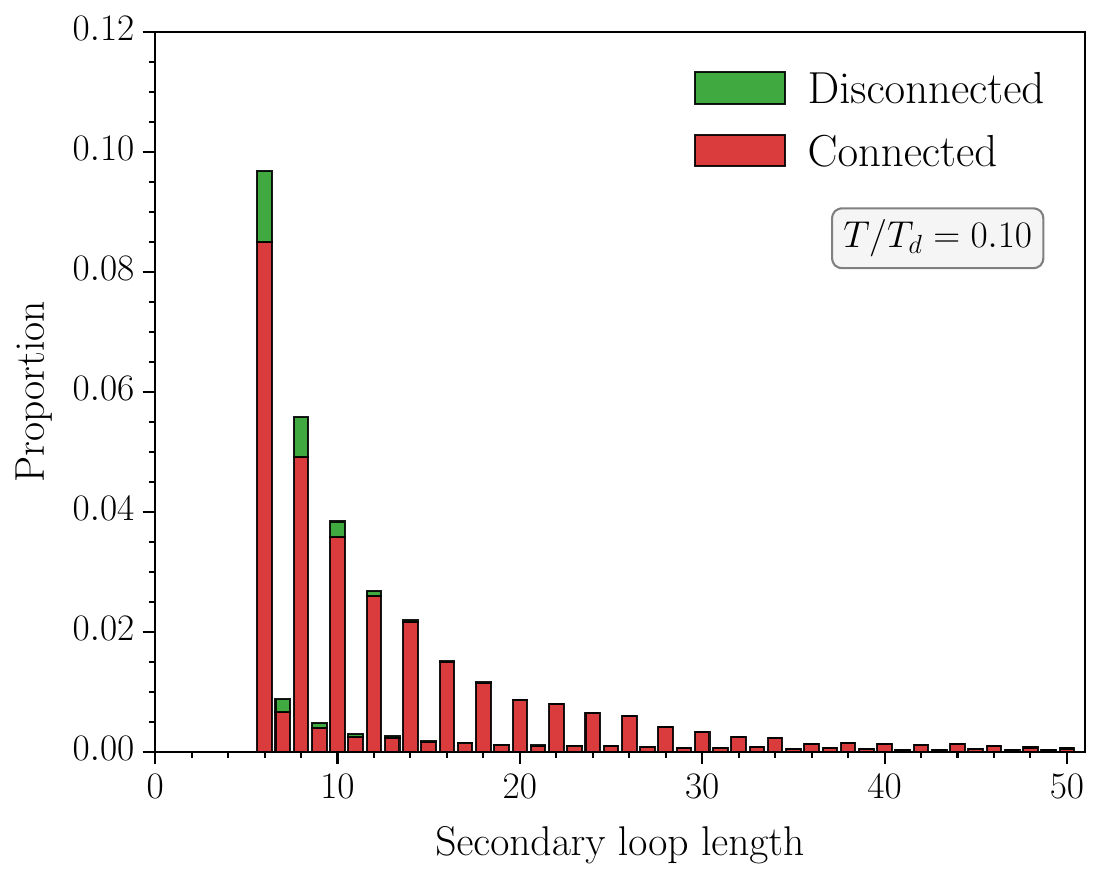}
	\hfill
	\includegraphics[width=0.48\linewidth]{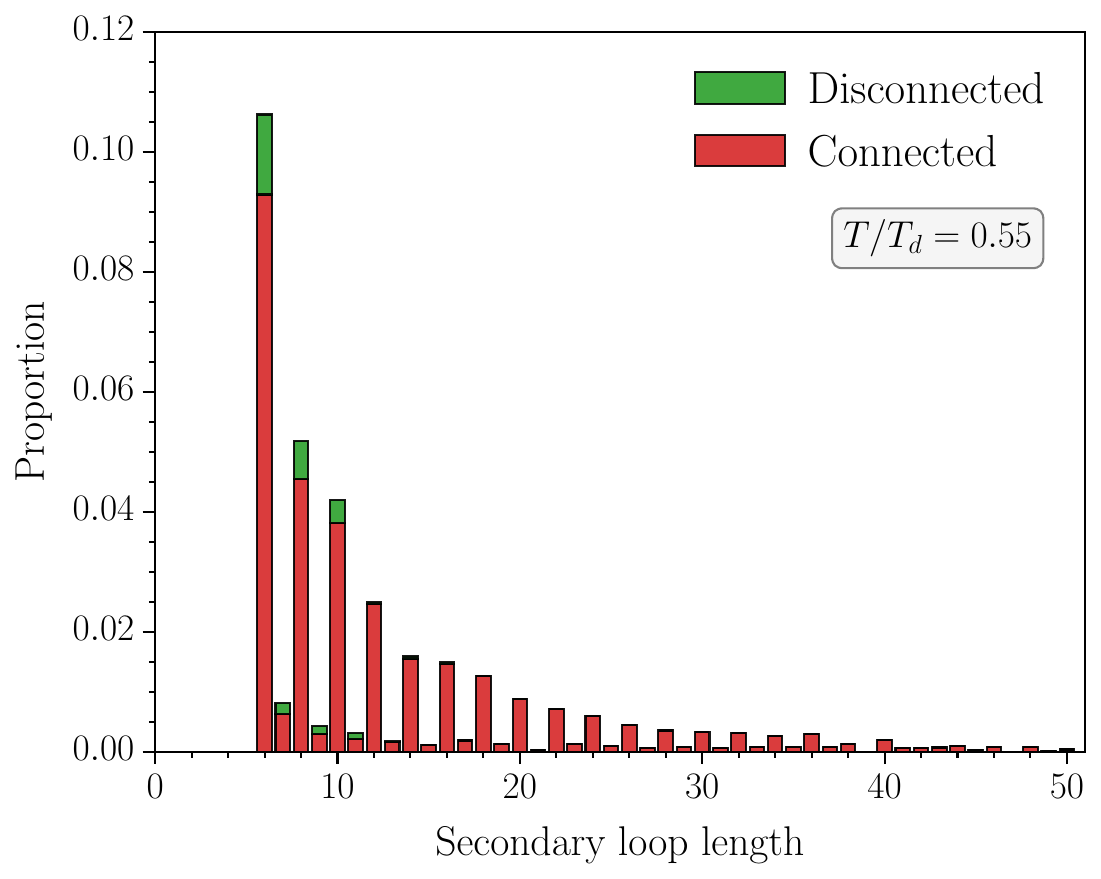}
	
	\vspace{1em}
	
	\includegraphics[width=0.48\linewidth]{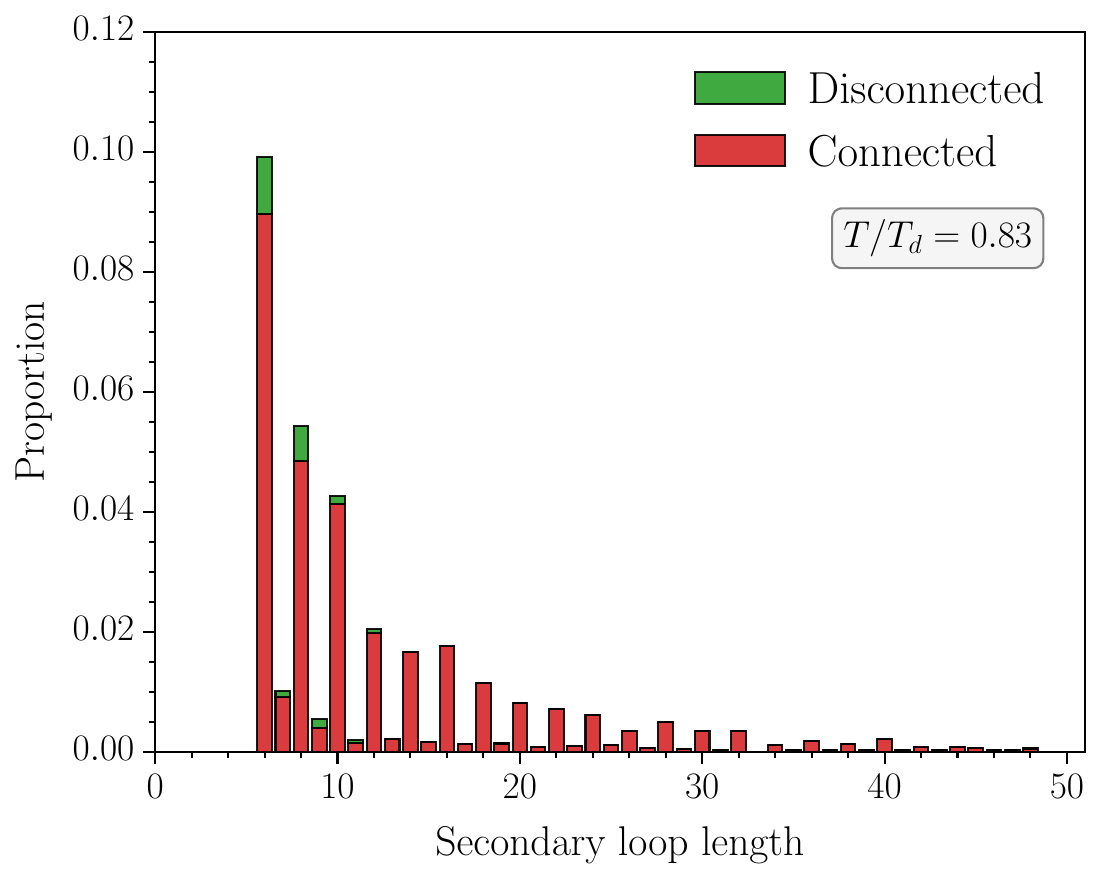}
	\hfill
	\includegraphics[width=0.48\linewidth]{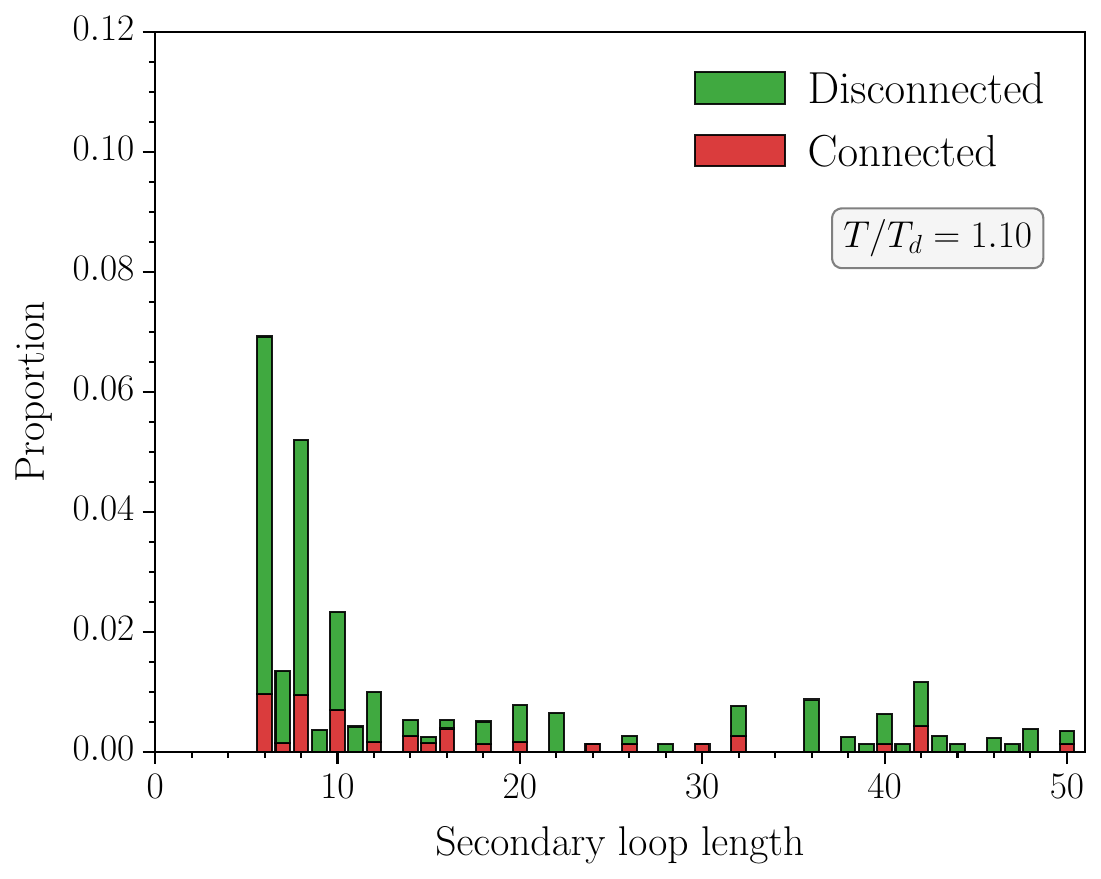}
	
	\vspace{1em}
	
	\includegraphics[width=0.48\linewidth]{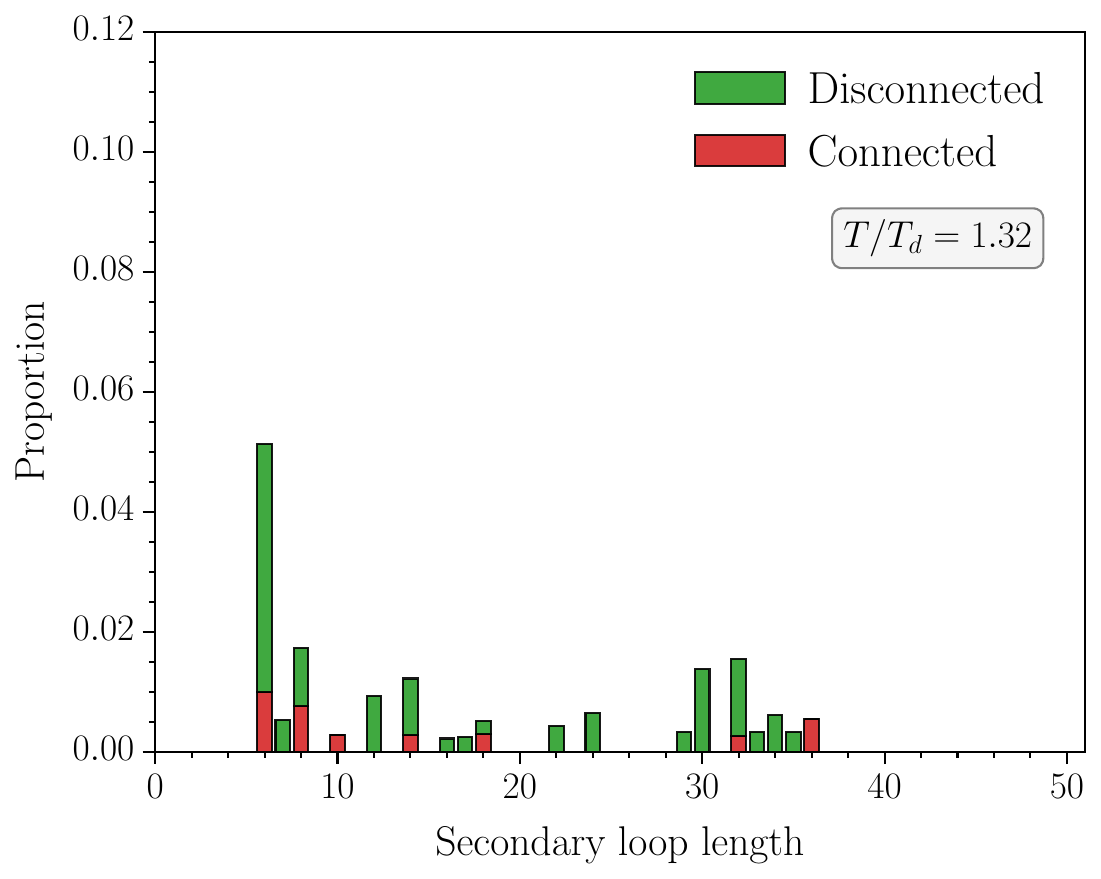}
	\hfill
	\includegraphics[width=0.48\linewidth]{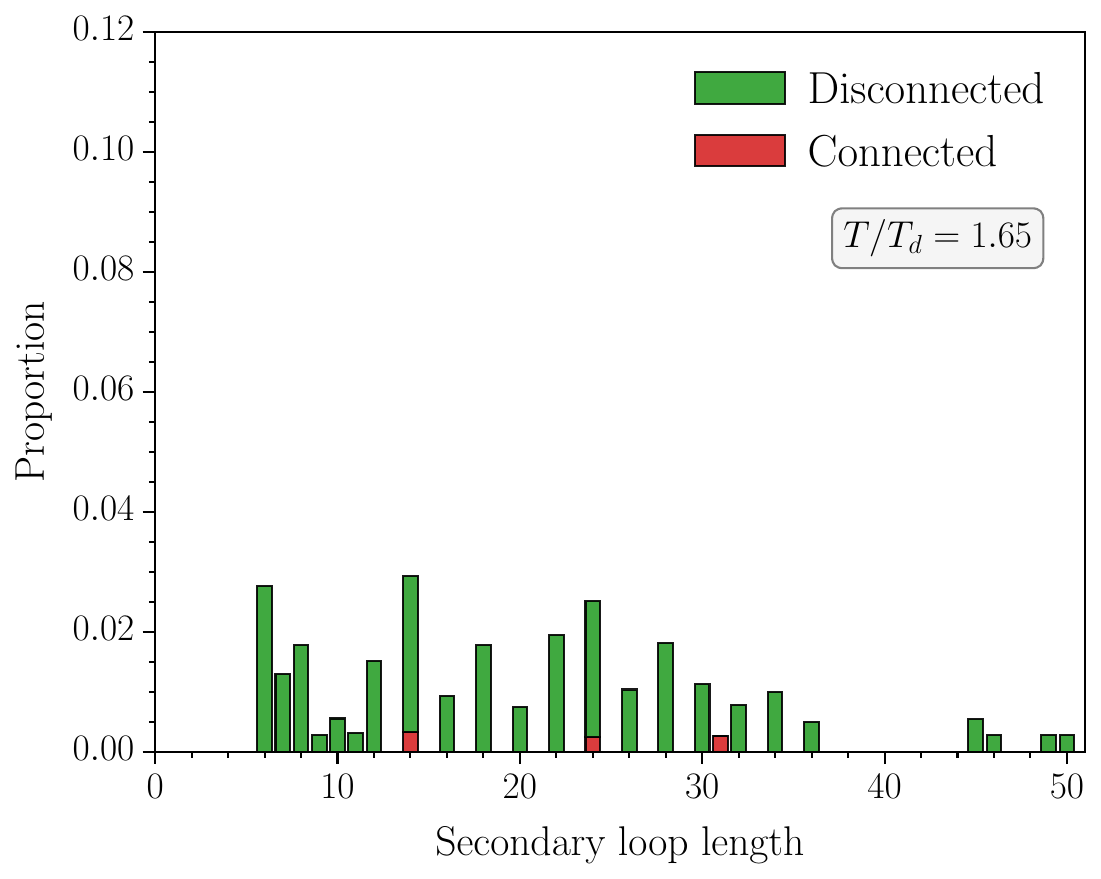}
	\caption{\label{fig:secondarylooplengthsexcludingfour} Same as in Fig.~\ref{fig:secondarylooplengths}, but excluding the elementary \(1 \times 1\) vortex loops of length four. The asymmetry between even and odd lengths is apparent at all temperatures, with the later suppressed due to requiring vortex branching.}
\end{figure*}

For completeness, the histograms are reproduced with elementary vortex loops of length four removed in Fig.~\ref{fig:secondarylooplengthsexcludingfour}.
This allows a closer inspection of longer secondary loops. As with the histograms of sheet sizes in Fig.~\ref{fig:secondarysheetsizes}, an asymmetry between even and odd loop lengths is evident. The latter are again suppressed, necessitating a branching point somewhere in the loop. Beyond lengths of \(\simeq 14\), the contribution of disconnected loops to the below-\(T_d\) histograms is no longer visible at the shown scale. This further emphasises how large secondary loops are almost certainly connected, arising from curvature in the primary sheet. We note that the longest disconnected secondary loop identified below \(T_d\), found at our lowest temperature, has a length of 23 vortices.

Above \(T_d\), there is insufficient statistics to produce a smooth histogram at larger loop lengths. This is from a combination of the smaller four-dimensional volume and the substantial reduction in number of secondary loops that occurs through the phase transition. Still, the important changes are apparent. Having removed the \(1 \times 1\) secondary loops, there are only a small handful of connected loops that contribute above \(T_d\). Instead, nearly all secondary loops are disconnected. The connected loop with a length of 24 that can be discerned in the highest-temperature histogram is the one shown in Fig.~\ref{fig:connectedloopaboveTd}.

\section{Conclusion} \label{sec:conclusion}
In this paper, the geometry of centre-vortex sheets in four dimensions was studied extensively in \(\mathrm{SU}(3)\) lattice gauge theory. By explicitly mapping out these sheets, a comprehensive picture of their relationship with secondary loops observed in three-dimensional slices of the lattice was developed, with an emphasis on the structural changes that occur through the phase transition.

Initially, this connection was examined qualitatively by producing visualisations of centre-vortex structure with two different colour schemes. In one scheme the vortex jets were coloured by the loop they belong to in the three-dimensional slice, and in the other by the sheet they belong to in four dimensions. This revealed that below \(T_d\), the majority of secondary loops that manifest in three-dimensional slices are connected in four dimensions, lying in the percolating sheet. These loops instead develop due to curvature in this sheet. Above \(T_d\), the short vortex lines that wind around the temporal dimension in spatial slices of the lattice were also identified to primarily lie in the same connected sheet, indicating that at high temperatures there is still a single large sheet that has simply oriented itself with the temporal dimension.

In addition to the primary sheet that exists at all temperatures, there are a handful of small secondary sheets scattered throughout the volume in four dimensions. A density of secondary sheets was computed and found to be approximately constant across the full temperature range with \(\rho_\mathrm{sheet} \simeq 1/3\)\thinspace fm\(^{-4}\). That is, the number of secondary sheets that form relative to the volume is independent of temperature. This is with the exception of a temporary increase that occurs just above \(T_d\), a recurring anomaly in most quantities considered in this work.

By producing histograms of secondary-sheet sizes, an asymmetry between even and odd sizes was observed, with the latter requiring a branching point to form a closed surface. Below \(T_d\), the secondary-sheet sizes plausibly obeyed a power-law distribution, though additional statistics are needed to verify this. Above \(T_d\), larger secondary sheets were found compared to below \(T_d\). This is possibly a consequence of the temporal alignment in the deconfined phase, with some secondary sheets also locked into winding around the temporal dimension and therefore persisting for the full temporal extent.

Thereafter, the properties of ``connected" (lie in the primary sheet) and ``disconnected" (lie in a secondary sheet) loops in temporal slices were thoroughly investigated. This quantitatively established that below \(T_d\), secondary loops tend to be connected and that disconnected loops represent only a minor contribution. Above \(T_d\), the average number of connected secondary loops in temporal slices fell to \(\simeq 0\), while the number of disconnected loops remained approximately constant with \(\simeq 1\) per temporal slice. The former behaviour is a direct consequence of the temporal alignment above \(T_d\), implying that the primary sheet almost never curves back on itself in the temporal dimension as is needed to produce connected secondary loops in temporal slices.

A small increase in connected secondary loops as \(T_d\) is approached from below was related to the spike in secondary-sheet density above \(T_d\). The extra connected loops must arise from additional curvature in the primary sheet. As the temporal alignment takes hold, these regions of curvature ``break off" from the sheet and form small secondary sheets just above \(T_d\). It will be interesting to perform these calculations at a larger spatial volume to explore aspects of the finite-volume crossover as one approaches a genuine phase transition.

Finally, the lengths of connected and disconnected secondary loops were examined. Below \(T_d\), connected loops form the dominant contribution at all lengths, including the \(1\times 1\) vortex loops of length four. Almost zero disconnected loops longer than \(\simeq 14\) vortices were present. Due to the suppression of connected loops at high temperatures, the reverse holds above \(T_d\). Here, any connected loops longer than four vortices are vanishingly rare.

Future work will apply the techniques developed here to the dynamical gauge field configurations of QCD. Centre-vortex geometry has previously been studied at finite temperature with dynamical fermions, revealing three distinct regimes for vortex behaviour \cite{Mickley:2024vkm}. The first two regions are separated by the chiral transition, while the second and third regions are separated by the percolation transition that is associated with loss of confinement. It will be interesting to see how the black-and-white findings of the pure-gauge theory extend to the more-complex centre-vortex behaviour with dynamical fermions.

\begin{acknowledgments}
It is a pleasure to thank Thakur Giriraj Hiranandani for discussions on analysing centre-vortex geometry in four dimensions. This work was supported with supercomputing resources provided by the Phoenix HPC service at the University of Adelaide. This research was undertaken with the assistance of resources and services from the National Computational Infrastructure (NCI), which is supported by the Australian Government. This research was supported by the Australian Research Council through Grant No.\ DP210103706. C.~A. is grateful for support via STFC Grant No.\ ST/X000648/1 and the award of a Southgate Fellowship from the University of Adelaide. R.~B. acknowledges support from a Science Foundation Ireland Frontiers for the Future Project award with Grant No.\ SFI-21/FFP-P/10186.
\end{acknowledgments}

\bibliography{main}

\end{document}